\documentclass[fleqn,usenatbib,letterpaper]{mnras}

\usepackage{newtxtext,newtxmath}

\usepackage[T1]{fontenc}
\usepackage{ae,aecompl}

\usepackage{graphicx}	
\usepackage{amsmath}	
\usepackage{amssymb}	
\usepackage{tabulary}
\usepackage{nccmath}    
\usepackage{enumitem}
\usepackage{bm}
\usepackage{comment}
\usepackage{array}
\usepackage{multirow}

\newcommand{\rowstyle}[1]{\gdef\currentrowstyle{#1}%
  #1\ignorespaces}

\newcolumntype{*}{>{\global\let\currentrowstyle\relax}}
\newcolumntype{^}{>{\currentrowstyle}}

\newcommand{\blank}{NP}
\newcommand{\orone}{OFP}
\newcommand{\ortwo}{NP/EB/OFP/WF}
\newcommand{\orthree}{NP/EB/OFP}
\newcommand{\nwnextgen}{NP/EB}
\newcommand{\nextgen}{NP/EB/WF}

\title[Classifying Exoplanet Candidates with CNNs]{Classifying Exoplanet Candidates with Convolutional Neural Networks: Application to the Next Generation Transit Survey}

\author[A. Chaushev \& L. Raynard]{
\parbox{\textwidth}{
Alexander Chaushev,$^{1}$\thanks{E-mail: a.chaushev@tu-berlin.de}
Liam Raynard,$^{2}$
Michael R.~Goad,$^{2}$
Philipp~Eigm\"uller,$^{3}$
David J. Armstrong,$^{4,5}$
Joshua T. Briegal,$^{6}$
Matthew R.~Burleigh,$^{2}$
Sarah L. Casewell,$^{2}$
Samuel Gill,$^{4,5}$
James~S.~Jenkins,$^{7,8}$
Louise D.~Nielsen,$^{9}$
Christopher A.~Watson,$^{10}$
Richard~G.~West,$^{4,5}$
Peter~J.~Wheatley,$^{4,5}$
St\'ephane~Udry,$^{9}$
Jose~I.~Vines$^{7}$
}
\\
\\
$^{1}$Center for Astronomy and Astrophysics, TU Berlin, Hardenbergstr. 36, D-10623 Berlin, Germany\\
$^{2}$Department of Physics and Astronomy, University of Leicester, University Road, Leicester, LE1 7RH, UK\\
$^{3}$Institute of Planetary Research, German Aerospace Center, Rutherfordstrasse 2, 12489 Berlin, Germany\\
$^{4}$Dept.\ of Physics, University of Warwick, Gibbet Hill Road, Coventry CV4 7AL, UK\\
$^{5}$Centre for Exoplanets and Habitability, University of Warwick, Gibbet Hill Road, Coventry CV4 7AL, UK\\
$^{6}$Astrophysics Group, Cavendish Laboratory, J.J. Thomson Avenue, Cambridge CB3 0HE, UK\\
$^{7}$Departamento de Astronom\'ia, Universidad de Chile, Casilla 36-D, Santiago, Chile\\
$^{8}$Centro de Astrof\'isica y Tecnolog\'ias Afines (CATA), Casilla 36-D, Santiago, Chile\\
$^{9}$Observatoire de Gen{\`e}ve, Universit{\'e} de Gen{\`e}ve, 51 Ch. des Maillettes, 1290 Sauverny, Switzerland\\
$^{10}$Astrophysics Research Centre, School of Mathematics and Physics, Queen's University Belfast, Belfast BT7 1NN, UK\\
}

\date{Accepted for publication 23 July 2019}

\pubyear{2019}

\begin{document}
\label{firstpage}
\pagerange{\pageref{firstpage}--\pageref{lastpage}}
\maketitle

\begin{abstract}

Vetting of exoplanet candidates in transit surveys is a manual process, which suffers from a large number of false positives and a lack of consistency.
Previous work has shown that Convolutional Neural Networks (CNN) provide an efficient solution to these problems. 
Here, we apply a CNN to classify planet candidates from the Next Generation Transit Survey (NGTS).
For training datasets we compare both real data with injected planetary transits and fully-simulated data, as well as how their different compositions affect network performance.
We show that fewer hand labelled lightcurves can be utilised, while still achieving competitive results.
With our best model, we achieve an AUC (area under the curve) score of $(95.6\pm{0.2})\%$ and an accuracy of $(88.5\pm{0.3})\%$ on our unseen test data, as well as $(76.5\pm{0.4})\%$ and $(74.6\pm{1.1})\%$ in comparison to our existing manual classifications. 
The neural network recovers 13 out of 14 confirmed planets observed by NGTS, with high probability.
We use simulated data to show that the overall network performance is resilient to mislabelling of the training dataset, a problem that might arise due to unidentified, low signal-to-noise transits.
Using a CNN, the time required for vetting can be reduced by half, while still recovering the vast majority of manually flagged candidates. 
In addition, we identify many new candidates with high probabilities which were not flagged by human vetters. 

\end{abstract}




\section{Introduction}
\label{sec:Intro} 

Exoplanets detected via the transit method constitute 80\% of the total confirmed population \citep{Akeson2013}\footnotemark. However, current detection methods produce large numbers of false positives. Since these candidates are analysed manually by several human vetters, this is a time consuming process that lacks consistency.

Recent results \citep{Shallue2018,Ansdell2018,Schanche2019,Osborn2019,Dattilo2019,Yu2019} have shown that a convolutional neural network (hereafter CNN) provides an efficient, automatic approach to classifying exoplanet candidates. 
A CNN can be used to reduce the time burden on human vetters, as well as to identify promising candidates which may have been missed, particularly those in lower signal-to-noise (S/N) regimes where there are many false positives.

\footnotetext{\url{https://exoplanetarchive.ipac.caltech.edu}, \mbox{online 15 September 2018}}

In this paper, we present the first application of a CNN to data from the Next Generation Transit Survey (NGTS) \citep{Wheatley2018} and show that it is effective in classifying exoplanet candidates found by \textsc{ORION}, an implementation of the Box Least-Squares detection algorithm (\textsc{BLS}) \citep{Collier-Cameron2006}.
We demonstrate that there is good agreement between the CNN ranking and our extensive database of classifications produced by expert human vetters. In addition, we build on previous work by investigating the optimal size and composition of the dataset used to train the neural network. 
Previous studies have relied on false-positive candidates, identified during the human vetting process, to formulate their CNN training data. By utilising transit injections we find that we can reduce the number of human labelled lightcurves needed for training, while still achieving competitive results. Labelling data is a time-intensive process and a key road block in training a CNN.

In \S\ref{sec:DatasetPreparation} we describe our datasets and data preparation procedures.
In \S\ref{sec:Network} we describe the architecture of the neural network and set out our methods for training and optimising the CNN.
We discuss the results of training using simulated data in \S\ref{sec:SimulatedData}. 
In \S\ref{sec:RealResults} we characterise the performance of our network  using real NGTS data. 
We draw comparison to human candidate classification in \S\ref{sec:EyeComparison} and describe our search for new, promising planet candidates in \S\ref{sec:NewCands}. Finally, we discuss our results and present our conclusions in \S\ref{sec:Discussion}.

\subsection{Transit search}
\label{sub:Transit_search}

Variants of the Box-Least Squares fitting (\textsc{BLS}) \citep{Kovacs2002} and matched filter \citep{Jenkins2002,Bord2007} methods have become the canonical tools for the detection of exoplanet signals in transit lightcurves. The facilities which use them include WASP \citep{Pollacco2006,Collier-Cameron2006,Collier-Cameron2007}, XO \citep{McCullough2005}, HATNet \citep{Bakos2007a}, CoRoT \citep{Cabrera2011}, Kepler \citep{Jenkins2010, Cabrera2012}, KELT \citep{Siverd2012,Kuhn2016}, MASCARA \citep{Talens2017}, NGTS \citep{Wheatley2018} and TESS \citep{Ricker2015}.
Unfortunately these methods yield vast numbers of false positives. 

For instance there are more than 58,500 targets with \textsc{ORION} candidates in NGTS data. With up to 5 different detections considered per target, this gives over 212,000 candidates in total. 
\citet{Gunther2017a} estimated that $\sim 97\%$ of these are false positives, reducing to 82\% after initial vetting tests.
These numbers are broadly consistent with false positives from other missions such as CoRoT and Kepler, which can range from $\sim 50\%$ to $\sim 90\%$ \citep{Deleuil2018, Akeson2013, Santerne2016}. 

Large numbers of candidates with a high false positive rate demand many resources during the vetting process, since candidates are analysed visually by a human being. 
It is also difficult to ensure consistency across expert vetters, as some of the judgements may be subjective, particularly for marginal candidates. 

\subsection{Deep Learning}

Machine learning is a subset of artificial intelligence which studies algorithms that `learn' to perform a task instead of following explicit steps.
Machine learning approaches have become increasingly popular in the field of exoplanet detection and vetting, and are being applied to address the shortcomings of transit detection algorithms.

\citet{McCauliff2015} and \citet{Mislis2016} utilised a random forest classifier \citep{Breiman2001} on Transit Crossing Events (TCEs) in Kepler data.
Others have used self-organising maps to group Kepler lightcurves with similar features and to classify new objects according to their similarity with each group \citep{Thompson2015,Armstrong2017}. 
\citet{Armstrong2018} combined a self-organising map with a random forest model to rank NGTS candidates produced by \textsc{ORION}.
The Autovetter algorithm achieved an Area Under the Curve (AUC) score of 97.6\% in ranking injected transits against false positives in the NGTS dataset.

More recently, a variety of machine learning techniques, called ``Deep Learning'', have provided performance improvements for many applications \citep{Lecun2015,khamparia2019}. 
A Deep Neural Network (DNN) consists of three or more layers of interconnected neurons, and is capable of learning useful features for classifying the data automatically. 
Performance of Deep Learning techniques have also been shown to scale well with large volumes of data \citep{Sun2017}.  
These traits are advantageous for the exoplanet candidate classification problem.
A CNN is a common form of a DNN, which is loosely based on the architecture of the animal visual cortex \citep{Lecun1990,Lecun1998}. 
CNNs are particularly suited to data that contains spatial structure, such as transit lightcurves when represented as one dimensional images.
Both \citet{Pearson2018} and \citet{Zucker2018} presented important case studies demonstrating the ability of CNNs to detect exoplanet candidates directly from lightcurves. 
However, they focused mainly on simulated data and did not proceed with applying their networks to search for new candidates. 

\citet{Shallue2018} applied their CNN, \textsc{AstroNet}, to classify new candidates in known planetary systems, using Kepler lightcurves. The authors showed that CNNs yielded greater success than alternative DNN architectures. 
A key result was that multiple ``views'' of the network input representation boosts performance.
\citet{Ansdell2018} built on this work by showing that incorporating the object centroid time series and stellar scalar properties (e.g. radius, temperature, density etc) in the network input, also increased the performance of the CNN. 
Other authors have applied CNNs to classify transiting exoplanet candidates: WASP: \citealt{Schanche2019}; K2: \citealt{Dattilo2019}; TESS: \citealt{Osborn2019,Yu2019}.

In this paper, we apply a CNN to classify exoplanet candidates in NGTS, developing a different method to that previously employed by \citet{Armstrong2018}. 
We draw a detailed comparison to human classifications from the vetting process and investigate how well the network performs with respect to the S/N of the transit detection. 

Importantly, we build on previous work by investigating how the composition of the non-planet class of the training dataset influences the network performance.
. 
We show that we can reduce the number of manually labelled lightcurves required for training, by utilising injections of planetary transits and astrophysical false positives instead, while achieving similar performance.
Finally, using simulated data, we show that network performance increases with training dataset size and also that it is robust to a small amount of contamination in the form of incorrectly labelled lightcurves.

\section{Dataset Preparation}
\label{sec:DatasetPreparation}

In this work we are concerned with distinguishing promising exoplanet candidates from false positives.
Therefore, we focus on training datasets with two classes; a planet class comprised solely of lightcurves with injected planetary transit signals and a non-planet class, containing either a false positive signal or no signal at all.
These are labelled as `1' and `0' respectively, with the CNN outputting a normalised score in this range that is interpreted as the probability of the lightcurve containing a transit. 

Previous studies utilised lightcurves of confirmed exoplanets and promising candidates identified from manual vetting, for their planet class \citep{Shallue2018,Ansdell2018,Dattilo2019,Yu2019}. Typically, the distribution of labelled lightcurves from the vetting process is highly imbalanced towards the non-planet class. 
Ideally the classes should be balanced so that the network is equally trained to recognise both.

Label imbalance is more prevalent in NGTS data, as the survey has not been operational for long enough to accrue a sufficiently large variety of confirmed planets and candidates, from which to produce a training set representative of the true population. 
Confirmed planets and planet candidates constitute only $1\%$ and $8\%$ respectively of NGTS vetting labels.
We note that even the long-established WASP survey was noted as having a deficiency of planet labels in their label distribution, with \citet{Schanche2019} opting to augment them via injection of simulated transits into real data. 
A similar strategy is necessary for the training of a CNN on NGTS data. 

Injection of artificial planetary transits into real data is a compromise which guarantees appropriate properties of the underlying data but with sufficient flexibility to produce transit signals of interest.
However, the use of lightcurves contaminated by real transit-like signals for either class, e.g. shallow signals, may confuse the network and lead to lower performance.
This potential issue was first highlighted by \citet{Zucker2018} and is discussed further in \citet{Yip2019}.

Fully simulated data is an alternative means of training a network, and one which offers full control over the parameter space as well as a pristine environment for validation and testing. 
The challenge for simulated data is in replicating the observation pattern and systematics inherent to the real data, such that the network is adequately trained for the task. 
\citet{Osborn2019} noted a reduced performance of their network when validated on real data, compared to \citet{Shallue2018, Ansdell2018}. The authors highlighted that training on simulated TESS data may be a contributing factor.
Indeed \citet{Yu2019} trained their network on real TESS data and achieved better performance, although we note that results from these studies are not directly comparable as there are many differences between the network inputs and the data themselves.

In this work we consider both fully simulated data and injections of simulated transits into real data, when training our CNN.
We discuss these separately in \S \ref{sub:RealData} and \S \ref{sub:SimulatedData}.
In addition, we also consider the effect of varying the dataset composition of the non-planet class.
We do this by including injections of artificial false positives such as eclipsing binaries, as well as true planet and false positive signals deliberately phase-folded on an incorrect period.
Previous studies concerned with the classification of real planet candidates, relied solely on the use of real false positive candidates identified via vetting. 

\subsection{NGTS Data}

\label{sub:RealData}

NGTS is a wide-field, ground-based transit survey located at ESO's Cerro Paranal observatory, Chile \citep{Wheatley2018}. 
NGTS comprises 12, fully roboticised 20 cm telescopes, each with an \mbox{8 deg$^2$} field of view. 
The goal of the NGTS project is to detect Super-Earths and Mini-Neptunes around bright host stars ($m_v <$ 13) suitable for radial velocity confirmation and atmospheric characterisation.
In survey mode, each NGTS field is observed for approximately 8 to 9 months, for periods of time starting at 30 minutes through to a full 8 hours of continuous coverage. 
An image is taken every 12 seconds with a 10 second exposure time. 
For a full discussion of the processing of NGTS data including reduction, photometry and detrending, the reader is referred to \citet{Wheatley2018}. 
Each NGTS lightcurve is further detrended to remove stellar noise and sidereal day artefacts using a custom built detrending pipeline (Eigm{\"u}ller in prep).
As part of the additional detrending, data points that are affected by bad weather or poor conditions are further removed from the lightcurves.
All data was drawn from the most recent NGTS pipeline run, which is called  `CYCLE1807\_DC' under the NGTS naming convention.

In total, 91 fields were available for processing with the neural network and from which data for a training set could be drawn. 
This comprises over 890,000 lightcurves brighter than $I_{NGTS}$ of 16th magnitude. 
While the primary goal of the survey is to find planets around bright host stars ($m_v <$ 13), all lightcurves down to 16th magnitude are searched.
Including these lightcurves increases the parameter space to which the neural network will be sensitive to and also allows us to boost our training dataset size.
Each lightcurve has on average 178,000 data points, up to a maximum of approximately 210,000.  
Six of the 91 fields have less than 100,000 measurements either due to weather, maintenance of equipment or ongoing observations for fields which are incomplete.

The \textsc{ORION} detection package \citep{Collier-Cameron2006} was run over the fully detrended data from all of the fields. 
\textsc{ORION} produced 212,000 candidate transit detections from 58,500 separate targets, with at least one detection having a signal detection efficiency (SDE) of greater than 5 (the lower threshold for the first detection).
\textsc{ORION} searches for candidates in the period range 0.35 to 35 days.
As part of routine NGTS operations, \textsc{ORION} candidates are regularly vetted by members of the consortium.
The vetting process is organised by observed field, with every NGTS field being vetted by at least two people. 
The initial screening of a field involves marking interesting candidates for discussion using a D flag. 
These D candidates are then further discussed by a larger group, before either being unflagged or labelled as AS, BS or AD if it is decided they are promising. 
False positives have their own flags and a full description of these can be seen in Table \ref{tab:ngts_flags}. 
Most candidates are left unlabelled if they are not likely to be real or do not conform to a clear false positive scenario.

\begin{table}
	\centering
	\caption{List of initial flags assigned by human ``eyeballers'' during the NGTS planet candidate vetting process. Promising candidates identified by individuals are first assigned a D flag, prompting discussion by the wider consortium. Following discussion, a different flag is assigned from one of two groups indicating whether the candidate requires further follow up or has been rejected as a false positive. If a candidate is subsequently confirmed as a planet, it is assigned a P flag.}
	\label{tab:ngts_flags}
	\begin{tabulary}{0.9\linewidth}{lL}
	{Flag} & {Description} \\
	\hline
	D & Marked for discussion \\ 
	\\
 	AD & Planet candidate with deep transit \\
	AS & Planet candidate with shallow transit \\
	BS & Planet candidate with shallow transit being held for further discussion before follow up \\
    \\
 	EA1 & One eclipse visible and otherwise flat \\
	EA2 & Two eclipses visible otherwise flat \\
	EB & Continuously variable but with contact points and/or V-shaped minima \\
	SINE & Sine-like continuously variable source (including asymmetric pulsators) \\
	OTH & Other variability \\
	UNF & Candidate was unflagged after further discussion \\
	\\
	P & Confirmed planet\\
    \hline
	\hline
	\end{tabulary}
\end{table}

\begin{table}
	\centering
	\caption{Allowed parameter ranges for injection of artificial planetary transits and eclipses of stellar binaries. The third light ratio is defined as the ratio of flux originating from a third body in the aperture, to the flux originating from the system of interest. Eclipsing binaries were injected into real data but not simulated data.}
	\label{tab:Transit_injection_params}
	\begin{tabulary}{0.9\linewidth}{LLL}
	{Parameter} & {Minimum value} & {Maximum value} \\
	\hline
 	Period & 0.1 days & 15.0 days\\
 	Duration & 10 min & 6 h\\
 	Third light ratio ($L_3$)& 0 & 1.0\\
 	\\
 
 	Planetary transits & & \\
 	Depth & 0.5 mmag & 6.0 \%\\
	$R_{planet}$ & 0.5 $R_{Earth}$ & 2.2 $R_{Jup}$\\
	$R_{star}$ & 0.2 $R_{\odot}$ & 2.0 $R_{\odot}$\\
 	$R_{planet}$ \big/ $R_{star}$ & 0.0022 & 0.25\\ \\
	
	Eclipsing binaries & & \\
	Depth & 0.5 mmag & 100 \%\\
	$T_{eff  A,B}$ & 3030 K & 9200 K \\
	$R_{A,B}$ & 0.2 $R_{\odot}$ & 2.0 $R_{\odot}$\\
 	$R_{B}$ \big/ $R_{A}$ & 0.1 & 10.0\\
	\hline
	\hline
	\end{tabulary}
\end{table}

To create the planet class of our network training, validation and test datasets, we first select a sample of lightcurves to be hosts for planetary transit injections by filtering out lightcurves with \textsc{ORION} candidates.
This reduces the likelihood that the remaining lightcurves contain real transits or false positive signals, which the network might confuse with the injected signals.
We utilised the \textsc{ELLC} package \citep{Maxted2016} to perform transit injections. Using a Monte Carlo method, parameters were drawn from allowed ranges set out in Table \ref{tab:Transit_injection_params}. Our goal was to produce the maximum variety of transit signals, sampled as uniformly as possible, and not to emulate real world distributions. For each injection, we first drew the following parameters uniformly from their allowed ranges: orbital period, transit depth, $R_{star}$ and  third light ratio ($L_3$). $L_3$ is defined as the ratio of flux from a third body in the aperture to that originating from the target of interest, and we fixed its value to 0 for 50\% of the time.
As can be seen from Table \ref{tab:Transit_injection_params}, our chosen range of periods for injections differs slightly from the \textsc{ORION} search period range (0.35-35 days). For the upper limit, detections of transits with periods greater than 15 days in the NGTS data are not very robust as often there is only a single transit event and a search in this regime would be more suited to a specialised effort. However, we note that in initial tests the CNN generalised well above this limit for those few \textsc{ORION} candidates in the long period regime, therefore we decided to include these in the comparison for completeness. The decision to the extend the lower limit for the injections was due to the fact that we may search this area in future. So we decided to choose a lower limit which was more physically justified than the ORION one.

To give our CNN a fair chance at detecting the transit signals, we ensured that the transit depth of any injected signal was no shallower than the standard deviation of the host light curve, when binned to 15 min cadence.
The planet-to-star surface brightness ratio and orbital eccentricity were both fixed to 0. 
Next we randomly chose to simulate either a full transit or a partial eclipse, each having equal probability. For the full eclipse regime, we numerically solved $R_p$ based on our choice of: transit depth, $R_{star}$ and $L_3$. Finally, we randomly selected an impact parameter in the range $0 < b \leq 1-k$, where $b$ is the impact parameter and $k$ is the planet-to-star radius ratio. For the partial eclipse regime, we instead numerically solved for the minimum allowed $R_p$ value, and then randomly selected a value for $R_p$ between this value and our maximum allowed limit in Table \ref{tab:Transit_injection_params}. Finally, we numerically solved for an impact parameter in the range $1-k < b \leq 1+k$.
Transit epochs were uniformly sampled in the range of 0 to the chosen orbital period, and the semi-major axis was set so as to permit the chosen transit depth. 

Valid injection signals were those which had a minimum of 3 transits, each covering at least one third of a transit, and where all parameters fell within the respective permitted ranges (Table \ref{tab:Transit_injection_params}).
We employed a simple trapezoidal transit model, neglecting the effects of limb darkening and the signal was strictly periodic (no transit timing variations). Similarly, we inject signals arising from a single planet per lightcurve, we did not consider multiplanetary systems. 

For the object centroiding time-series we applied shifts to the measured CCD x- and y-position values, co-incident with transit events in the flux time series, with 50\% probability. 
When applied, the shifts were proportional to the flux dilution parameter with a maximum absolute value of 0.5 pixels.
Injecting the transit directly into the time-series in this way proved to be equivalent to simulations of the centroid shift done directly using a pixel level simulation.

For the non-planet class of the training, validation and test datasets, we consider four categories of false positives. These are:
\begin{nobreak}
\begin{itemize}
    \item \textbf{`Non-periodic (NP)'} - lightcurves with no \textsc{ORION} candidates, i.e. they contain no easily detectable, periodic transit-like signals
    \item \textbf{`Eclipsing binary' (EB)} - Non-periodic lightcurves with injections of eclipsing binary signals
    \item \textbf{`Wrong fold' (WF)} - planetary transits and eclipsing binaries folded on a randomly selected wrong period
    \item \textbf{`\textsc{ORION} False positive' (OFP)} - \textsc{ORION} candidates rejected as false positives during the vetting process
\end{itemize}
\end{nobreak}

For the EB category, we inject artificial binary eclipses into host lightcurves with no \textsc{ORION} candidates, in a similar way to planetary transit injections. However, for EBs we also include stellar effective temperatures ($T_{eff}$) as injection parameters, in addition to orbital period, eclipse depth, $R_{star}$ and $L_3$. These are uniformly sampled within the limits set out in Table \ref{tab:Transit_injection_params}. The stellar surface brightness ratio of the two components is then considered in the eclipse model.

OFP lightcurves are drawn from the pool of \textsc{ORION} candidates which received either one of the following flags during the vetting process: EA1, EA2, UNF, SINE, OTH, or received no specific flag.
These false positives have been checked by at least two independent vetters who both decided the candidate was not worth discussing further, as they were confident it was unlikely to be of a planetary nature.  
Those OFPs without flags include many targets with lower SDEs, whose true nature is less certain. 
It could be argued that potential real planetary signals are being introduced into the non-planet class data in this way. 
While we do expect that some good candidates may have been missed in the eyeballing process, the vast majority of these are expected to be false positives, up to 97\% as estimated by \citet{Gunther2017b}.
We directly investigate the effects of signal contamination in \S \ref{sec:SimulatedData}. 

\begin{figure}
	\centering
	\includegraphics[width=0.45\textwidth]{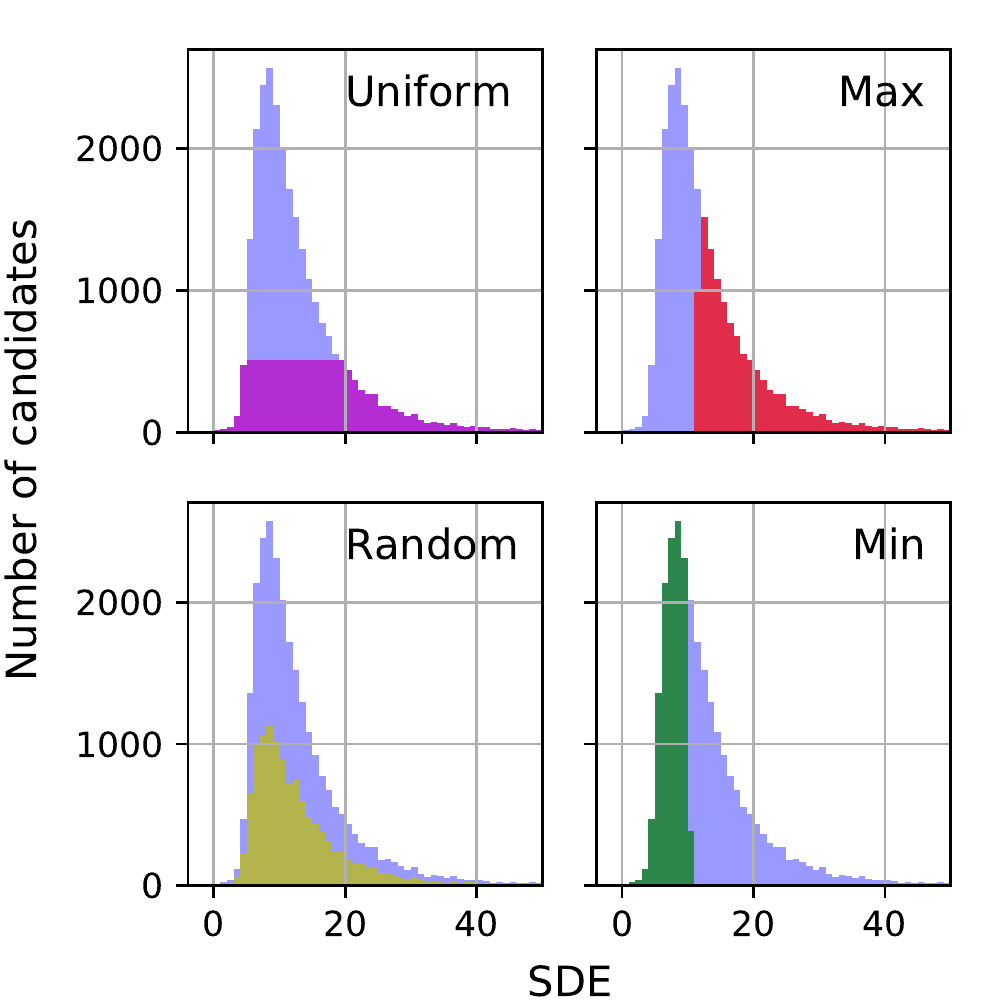}
    \caption{Distribution of SDE values for \textsc{ORION} false positive candidates from 45 fields, corresponding to half the dataset, are indicated by blue bins. Each of the four subplots shows a sample which can be selected using different criteria: uniform (purple bins), max (red bins), random (yellow bins) and min (green bins). By including each sample in turn for the OFP category, we investigate how the selection method affects network performance.}
    \label{fig:FPDist}
\end{figure}

\textsc{ORION} false positives have a broad range of SDEs, as such there are multiple ways of selecting them for inclusion in the negative class. Previous studies utilised the entire pool of false positives for training. 
However, the SDE distribution of false positives may influence the network's sensitivity to low and high signal-to-noise candidates. 
Therefore, we investigated how the mean SDE of the non-planet class affects the final network performance.
We consider false positives drawn in four different ways, as shown in Fig. \ref{fig:FPDist}, representing: a randomly drawn sample, a sample of the lowest SDEs, a sample of the highest SDEs and a sample drawn uniformly across the SDE range. 

In total we produced fifteen different datasets, differing in the composition of their negative class. 
There are six unique combinations of subclasses, each containing up to a maximum of four subclasses. 
The datasets which contain the \textsc{ORION} false positive subclass each have four variants, in which lightcurves were drawn from the SDE distribution in different ways (Fig. \ref{fig:FPDist}). 
Each of the fifteen dataset compositions contain 24,000 lightcurves in the training dataset, split evenly between the transit and non-planet class.
Where there are two or more subclasses, each subclass contains an equal number of lightcurves. 
A summary of the different datasets are given in Table \ref{tab:dataset_summary}.

An independent test of network performance must be carried out using previously unseen data. We aim to classify \textsc{ORION} candidates, however our training datasets with the false positive subclass contain a sample of the same lightcurves.
To avoid training and evaluating on the same lightcurves, we divide the \textsc{ORION} candidates into two groups according to NGTS field.
Fields with an RA of less than 12 hours comprise the first group, while fields with RA of 12 hours or more make up the second group. 
For each dataset with the OFP subclass, we train two separate versions of the network, one for each group. 
Network performance for group one is evaluated on group two and vice versa.

\begin{table}
	\centering
	\caption{Summary of the different neural network training datasets used in this study. Values indicate the number of lightcurves each training dataset comprises, in units of one thousand lightcurves, broken down by class and sub-class. Models trained on simulated data and real data are grouped separately. The planet class is composed of synthetic planetary transits injected into either real or simulated lightcurves. The non-planet class is composed of up to four sub-classes: non-periodic (NP), eclipsing binary (EB), wrong fold (WF) and \textsc{ORION} false positives (OFP), which are defined in \S \ref{sub:RealData}. The OFP selection refers to one of four distributions use to select the \textsc{ORION} false positives via their SDE. These are shown in Fig. \ref{fig:FPDist}.}
	\label{tab:dataset_summary}
 
    \begin{tabulary}{\linewidth}{ll|RRRr|C}
	\multirow{2}{*}{Model Name} & {OFP} & \multicolumn{4}{c|}{Non-Planet class} & {Planet} \\ \cline{3-6}
    & {Selection} & {NP} & {EB} & {WF} & {OFP} & {Class} \\
	\hline
	{Simulated Data:} & & & & & & \\
	- & - & 12 & -  & - & - & 12 \\
 	- & - & 50 & - & - & - & 50 \\
    {} & {} & {} & {} & {} & {}  & {} \\
	{Real Data:} & & & & & &\\
	\blank & - & 12 &  - & -  & - & 12 \\
	\nwnextgen  & - & 6 & 6 & - & - & 12 \\
    \nextgen & - & 4 & 4 & 4 & - & 12 \\
    \ortwo & Max & 3 & 3 & 3 & 3 & 12 \\
     & Min & & & & &  \\
     & Random &  &  &  &  &  \\
     & Uniform &  & &  &  &  \\
    \orthree & Max & 4 & 4 & - & 4 & 12 \\
     & Min &  &  &  & &  \\
     & Random &  &  &  &  &  \\
     & Uniform &  &  &  &  &  \\
     \orone & Max & - & - & - & 12 & 12 \\
     & Min &  &  &  &  &  \\
     & Random &  &  &  &  &  \\
     & Uniform &  &  &  &  &  \\
     
    \hline
    \hline
	\end{tabulary}
\end{table}

\subsection{Simulated Data}
\label{sub:SimulatedData}

We generated 100,000 pure noise light curves that modelled the observational properties of the NGTS survey.
To determine the time sampling of each light curve we first defined a corresponding pseudo-field. 
For each field we chose the baseline length of night from a uniform distribution in the range of 7 to 9 hours.
We modelled the duration of darkness at Cerro Paranal between astronomical dusk and dawn, during the course of a year, with a sinusoid function and chose a random phase corresponding to the epoch at which observations commenced.
Beginning with a rising field visible for 30 mins at the end of the first night, and which rises 4 mins earlier each successive night, we stepped through nights to construct a time series with the maximum length of night set by the chosen baseline.

Each night we sampled the observation window every 10 mins up to either a total of 4,278 data points or when the field became visible for less than 30 mins, whichever came first. 
We added noise in the form of time offsets by drawing both the nightly observation start times and durations from normal distributions, with means equal to their nominal values and standard deviations of 10 mins. 
To simulate bad weather and operational issues, we implemented entire night drop outs with a probability of 35\% and intra-night drop outs of a random number of adjacent points with probability 5\%. 
To obtain the corresponding flux to the light curve time series, we used the Gaussian Process (GP) kernel from \citet{Zucker2018} to simulate intrinsic stellar variability with quasi-periodic and white noise components:
\begin{equation}
\label{eq:GPkernel}
\begin{aligned}
k(t_i,t_j) = &  A_{s}^{2}\;exp\Bigg[-\bigg(\frac{t_i-t_j}{\lambda_s}\bigg)^2\Bigg]\\
& +A_{q}^2\;exp\Bigg[-\frac{1}{2}sin^2\bigg(\frac{\pi(t_i-t_j)}{T_q}\bigg)-\bigg(\frac{t_i-t_j}{\lambda_q}\bigg)^2\Bigg]\\
&+A_{w}^2\delta(t_i-t_j) \; ,
\end{aligned}
\end{equation}

\noindent where $A_s$, $A_q$ and $A_w$ are the amplitudes of each component; $\lambda_s$ and $\lambda_q$ are the length scales of variations in the time axis; $T_q$ is the period of the periodic component; $t_i$ and $t_j$ are the times at different epoch and $\delta$ is the Kronecker delta. 
We implemented our GP kernel using the \textsc{GEORGE} package \citep{Ambikasaran2015}. 
The hyperparameters of the kernel were drawn from uniform distributions within the limits set out in Table \ref{tab:GP_kernel_hyperparams}. 
We utilised the same periodic limits as \citet{Zucker2018} but we chose a range of amplitudes spanning larger values, as NGTS is not as sensitive as Kepler. 
For each light curve we selected a corresponding stellar radius and V-band magnitude from uniform distributions in the range \mbox{0.2 $R_{\odot} \leq$ R $\leq$ 2.0 $R_{\odot}$} and \mbox{8 mag $\leq$ V $\leq$ 16 mag}, respectively. 
We utilised the following relation for the white noise amplitude and V-band magnitude:

\begin{equation}
\label{eq:Aw_V}
  A_w = a\;exp\bigg[\frac{0.4\big(V-8.0\big)}{b}\bigg]  \; .
\end{equation}

Parameters $a$ and $b$ in Eq. \ref{eq:Aw_V} were determined by fitting to the NGTS noise model from \citet{Wheatley2018}, giving 61 $\mu$mag and 0.59 $\mu$mag, respectively. 

To emulate real data artefacts, we created outliers by re-scaling randomly chosen flux points with an occurrence probability and maximum adjustment of 1\%. 
Simulated stellar flares were also injected using the model from \citet{Davenport2014}, with occurrence probability of 5\%. 
We chose the flare amplitude and duration uniformly from the ranges 0 to 7\% and 20 to 75 mins, respectively. 

We note that our chosen cadence of 10 mins is much longer than the actual 12 second cadence of the NGTS survey, and was a practical compromise since the time taken to sample from the GP scaled as the number of points to the third power. 
The effect of increasing the cadence is analogous to binning up the data, since for NGTS data white noise dominates the light curves on these time scales. 

\begin{table}
	\centering
	\caption{GP kernel hyperparameter ranges used in Eq. \ref{eq:GPkernel}, from which values are sampled in order to create fully simulated datasets. A full explanation of these datasets is given in \S \ref{sub:SimulatedData}.}
	\label{tab:GP_kernel_hyperparams}
	\begin{tabulary}{0.9\linewidth}{LLL}
	{Hyperparameter} & {Minimum value} & {Maximum value} \\
	\hline
 	$A_s$ & 200 $\mu$ mag & 5 mmag\\
 	$A_q$ & 200 $\mu$ mag & 5 mmag\\
 	$\lambda_s$ & 1 min & 10 h\\
 	$\lambda_q$ & 1000 min & 500 h\\
 	$T_q$ & 10 h & 500 h\\
	\hline
	\hline
	\end{tabulary}
\end{table}

To create network training, validation and test datasets, we formulate the planet class by injecting artificial planetary transits into half of the 100,000 lightcurves using the same procedure as for real data, described in \S\ref{sub:RealData}. 
For the non-planet class we take the remaining 50,000 light curves with no modifications.

\subsection{Input Representations}
\label{sub:Input Representations}

\begin{figure}
	\centering
	\includegraphics[width=0.8\columnwidth]{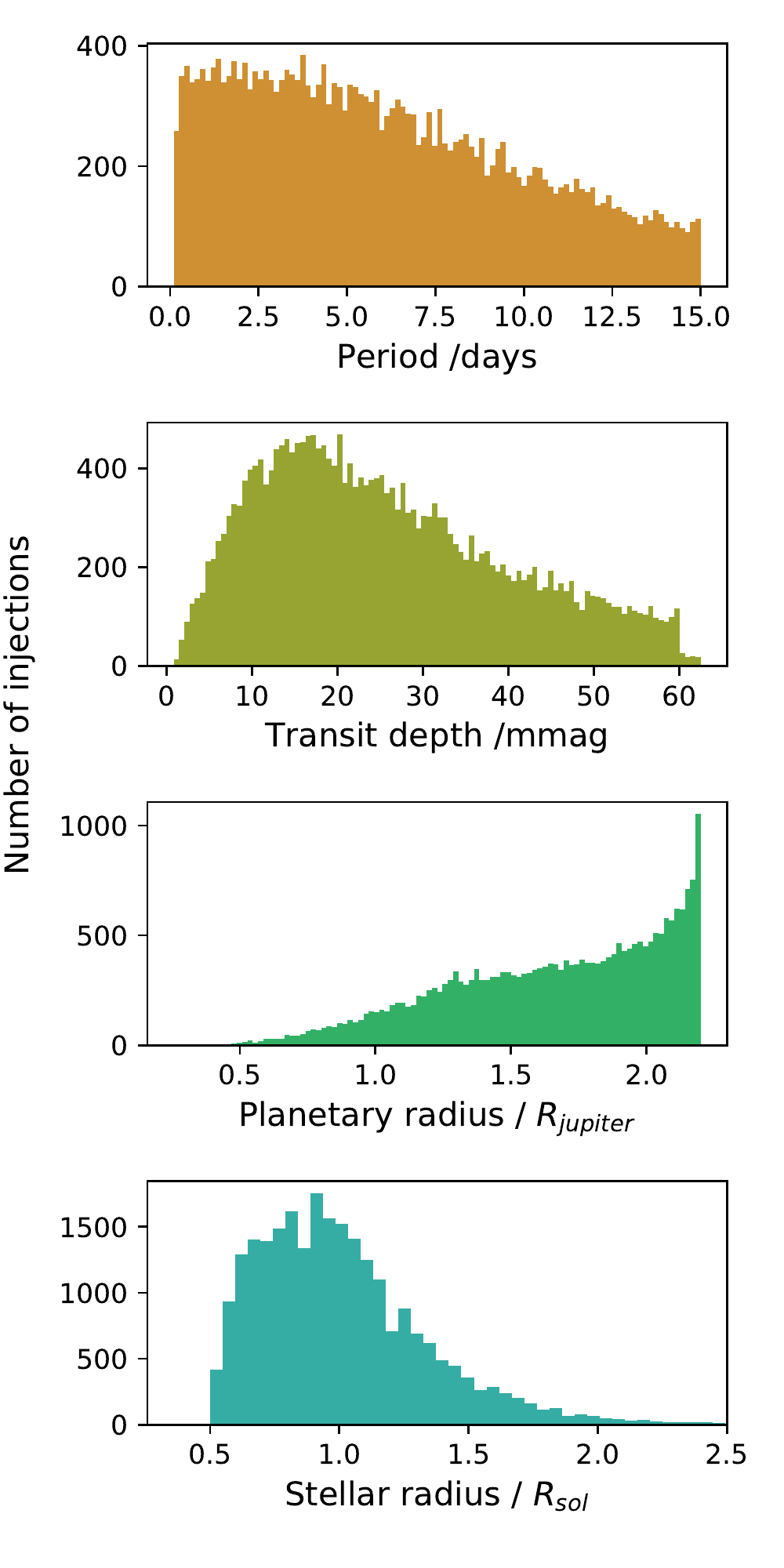}
    \caption{Posterior density distributions of transit injection parameters, for the planet class of real datasets.
    Although period, transit depth and stellar radius parameters are originally sampled uniformly, our Monte Carlo approach combined with our allowed parameter combination criteria result in departures from uniformity. For partial eclipses, the planetary radius is uniformly sampled. However, for full transits and eclipses the planetary radius is numerically solved based upon the chosen transit depth, stellar radius, surface brightness ratio and third light ratio. The distribution of planetary radii is skewed towards larger values since for partial eclipses, larger radii can produce the same transit depth as a smaller planet undergoing full transit, if the impact parameter is increased proportionately.}
    \label{fig:InjectedDistributions}
\end{figure}

\citet{Shallue2018} utilised both ``global'' and ``local'' views of their lightcurves, covering the entire lightcurves and a limited region of the primary transit event, respectively. 
They found that while the global view shows the out-of-transit noise as well as any secondary eclipses, the local primary view draws out the details of the primary transit. 
This is particularly important for short duration transits and longer orbital periods. 
We adopted this method but expanded it to include local views of any secondary transit, as well as the primary event. 
\citet{Ansdell2018} incorporated auxiliary scalar stellar host properties, as well as the target centroid time series in their network input representations. The former allowed the network to discriminate transit-like signals from signals consistent with exoplanet transits. 
The latter allows identification of centroid shifts indicative of diluted binary star eclipses, a common false positive. 
We adopted the centroid views and auxiliary stellar properties as inputs to our network. 

First, we generated global view input representations of the entire lightcurve flux series. We phase folded the lightcurves on their orbital periods, ignoring transit epoch, such that the transit event can be centred at any phase value. 
This makes the network more robust to uncertainties in ephemerides for \textsc{ORION} candidates and improved performance during early tests.
Bad datapoints, such as those with non-zero flags from the pipeline output, were removed. The lightcurves were then split into the same number of uniformly spaced bins. 
We normalised the lightcurve views such that the maximum depth had a value of -1 and the median (baseline) value was 0.

The global views of the centroid series were generated in the same way as per the flux, except that we did not normalise by the maximum depth. 
Instead, following \citet{Ansdell2018}, we normalised by the standard deviation of the centroid series scaled by that of the flux series, calculated from the out-of-transit regions across the entire dataset.

Local views of the flux and centroid series were produced in a similar way to the global views, but instead of using the whole lightcurves, we considered windows of the phase 0 and 0.5 regions, spanning 3 times the average transit duration of the confirmed exoplanet population (3.23 hours). 
To account for uncertainties in transit ephemerides in a similar way to the global views, we randomly offset the events from the centre of the window, up to a maximum of $2/3$ of the centre to edge span.

We opted to provide the orbital period as an auxiliary scalar input, to explore whether it can be utilised by the network to disregard spurious signals resulting from the observation window function of the NGTS survey, e.g. signals whose periods are integers of a day, or harmonics. 
Secondly, normalising the maximum depth of the flux series views allows the network to better interpret the data. However, in doing so we destroy information about the real transit depth. 
To prevent this information from being lost, we provide the maximum depth normalisation factor as an auxiliary input. 
Finally, the stellar host radius was included to allow discrimination between real exoplanet and exoplanet-like transits. 
For example, a deep transit of a large star is more likely to be an eclipsing binary system rather than an exoplanet transit.
The auxiliary scalar inputs were normalised by the standard deviations of their respective distributions.

For network training and evaluation, ideally the distributions of lightcurve transit injection and stellar host properties would be uniform, since we consider a broad range of planet, stellar host and lightcurve properties, as shown in Fig. \ref{fig:InjectedDistributions}. 
Although parameters were initially sampled from uniform distributions, their non-linear relationships coupled with a Monte Carlo selection method result in departures from uniformity. 
Balancing the value distributions of multiple parameters in combination is a non-trivial task.
In addition, lightcurves belonging to the non-periodic subclass did not undergo transit injection and so were not assigned transit related parameters.
For these, we sampled ephemerides and auxiliary stellar scalar values from the planet class population, so as not to introduce any biases in the training procedure.

\begin{figure*}
	\centering
	\includegraphics[width=0.98\textwidth]{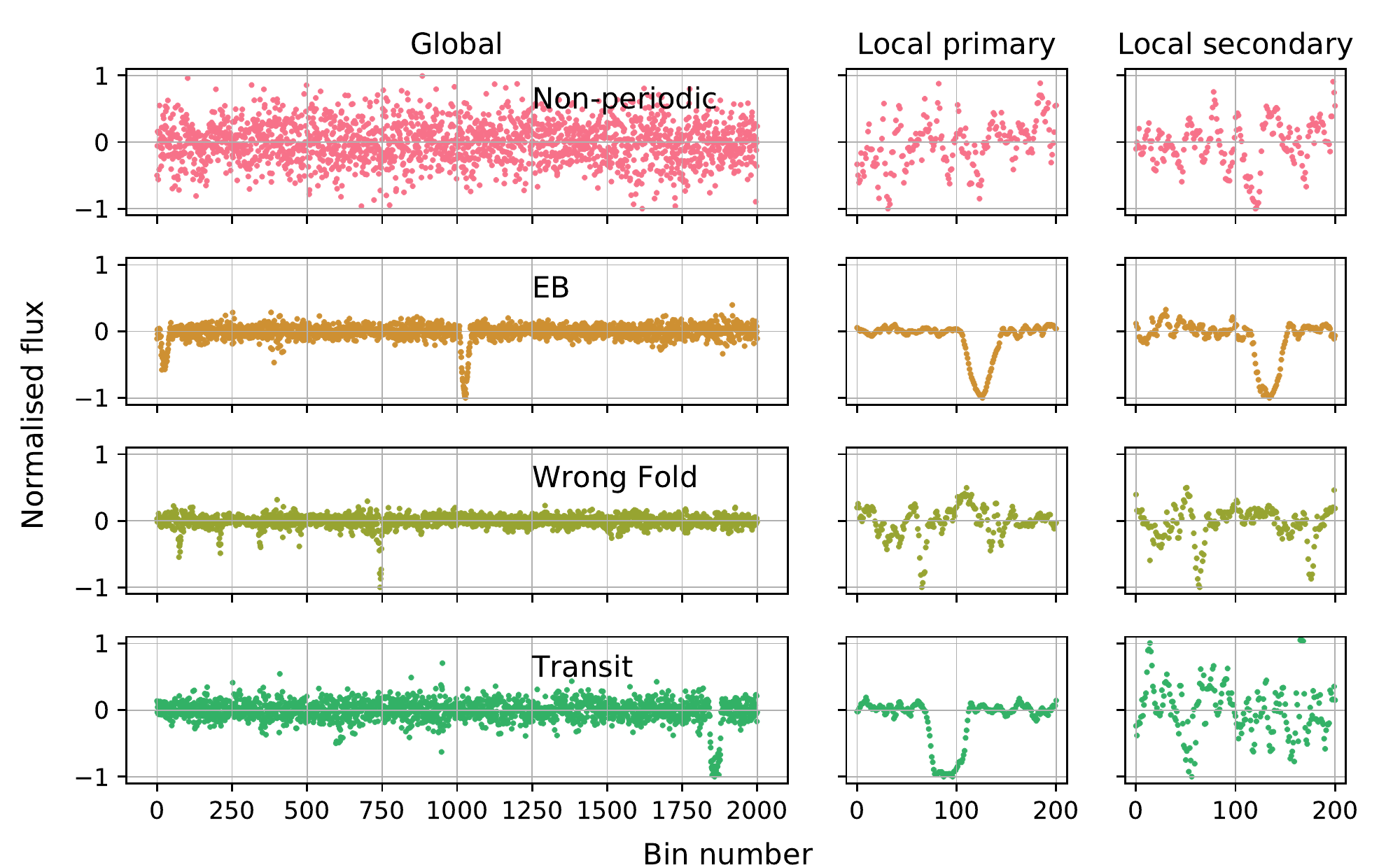}
    \caption{Example global and local view inputs for the phase folded lightcurves. The top three rows show three of the four categories of non-planet class lightcurves: Non-periodic (NP), eclipsing binary (EB) and wrong fold (WF); \textsc{ORION} false-positives (OFPs) are not shown. The bottom row shows an example lightcurve from the planet class.
    Lightcurves has been normalised to have a median value of 0 and maximum depth of -1. To account for uncertainties in \textsc{ORION} ephemeris, transit epoch is ignored for global views when phase folding the lightcurves, so the transit event can have any phase. Similarly for local views, the transit event is deliberately offset from the window centre.}
    \label{fig:InputRepExample}
\end{figure*}

In summary, we generated the following input representations:
\begin{itemize}
    \item Global view of flux series
    \item Global view of centroid series
    \item Local primary transit view of flux series
    \item Local primary transit view of centroid series
    \item Local secondary transit view of flux series
    \item Local secondary transit view of centroid series
    \item Auxiliary scalar orbital period
    \item Auxiliary scalar depth normalisation factor
    \item Auxiliary scalar stellar host radius
\end{itemize}
Fig.~\ref{fig:InputRepExample} depicts the flux input representations for the planet class and for 3 of the 4 subclasses of the non-planet class.

\section{Neural Network Architecture and Training}
\label{sec:Network}

\begin{figure}
	\centering
	\includegraphics[width=0.9\columnwidth, trim={8.5cm 0 0 0}, clip]{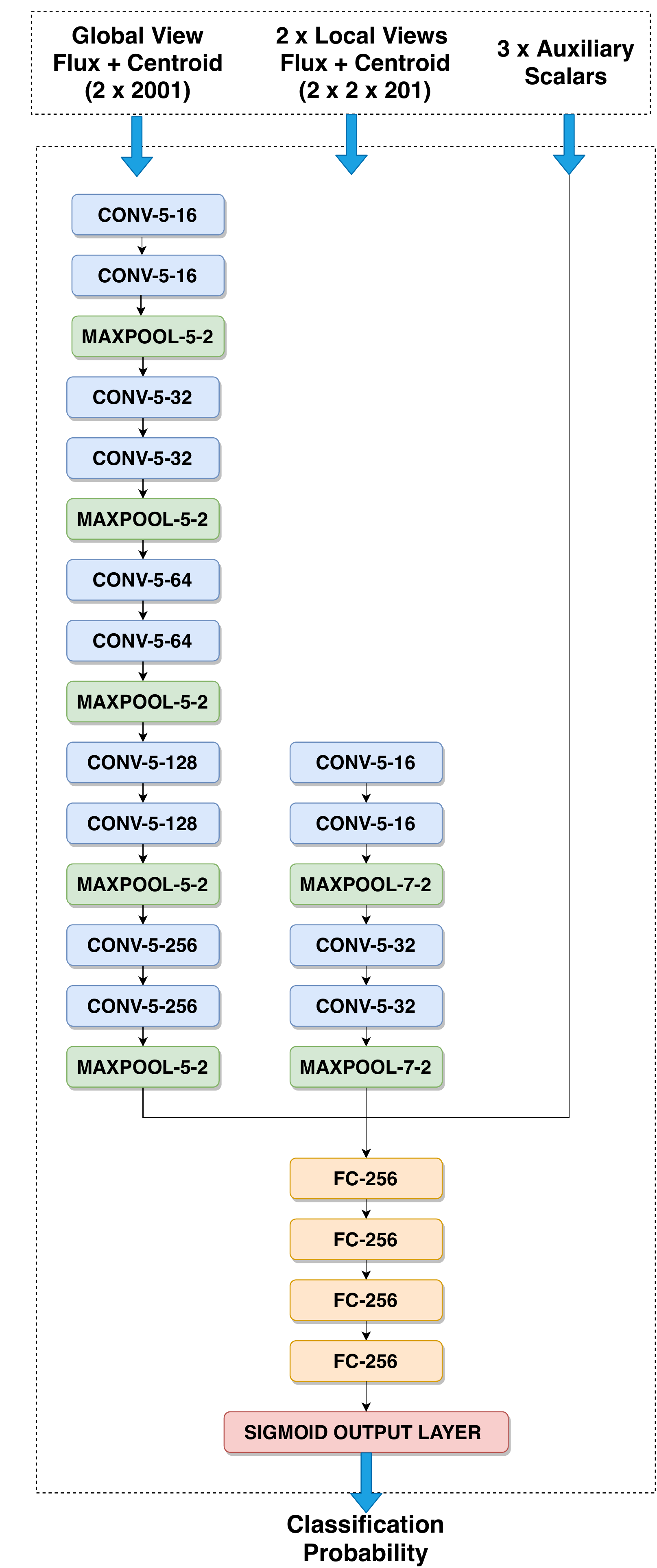}
    \caption{Architecture of our best CNN model. 
    Network inputs are passed through repeated blocks of convolutional and max pooling layers; global views, local views and auxiliary scalars are stacked respectively and passed through adjacent columns.
    The outputs from the different columns are combined prior to being passed through fully-connected layers. Convolutional layers are denoted Conv-\{kernel size\}-\{number of feature maps\}, max pooling layers are denoted MAXPOOL-\{window length\}-\{stride length\} and the fully connected layers are denoted FC-\{number of units\}. The output of the final, sigmoid layer is the predicted probability that each lightcurve contains a transiting exoplanet.}
    \label{fig:CNNArchitecture}
\end{figure}

Fig. \ref{fig:CNNArchitecture} shows the structure of the CNN used in this work. 
The architecture has been adopted from \textsc{AstroNet} \citep{Shallue2018}, including the use of a global and local view, and all parameters governing the fully-connected, pooling and convolutional layers.
However, we extended it by utilising additional inputs from other work \citep{Ansdell2018,Osborn2019,Yu2019} as discussed in \S \ref{sub:Input Representations}.
Our neural network is called `PlaNET' and was implemented using the \textsc{PyTorch} Python package \citep{Paszke2017}.

\subsection{Optimisation} 

\begin{table}
	\centering
	\caption{Hyperparameters and corresponding trial values used in our search for the optimal neural network architecture and training method. We abbreviate the following terms: Global View (GV), Local View (LV), Max Pooling (MP) and Fully Connected (FC). For the GV and LV, we define a block of layers as 2 convolutional layers followed by a MP layer.}
	\label{tab:optimisation_params}
	\begin{tabulary}{0.9\linewidth}{LL}
	{Hyperparameter} & {Trial values}\\
	\hline
	No. training epochs & 5, 10, 15, 20, 25, 30, 40, 50\\
 	\textsc{ADAM} learning rate & [5.0E-6, 1.5E-5]\\
 	Dropout probability & 0, 0.125, 0.25, 0.375, 0.5\\
 	GV kernel size & 3, 5\\
 	No. layers in block for GV & 1, 2\\
 	No. blocks of layers for GV & 1, 2, 3, 4, 5, 6\\
 	Conv. filter size in GV & 2, 4, 6, 8, 16\\
 	MP layer kernel size for GV & 3, 5\\
 	MP layer stride length for GV & 1, 2, 3\\
 	GV input vector size & 1001, 2001, 3001\\
 	LV kernel size & 3, 5\\
 	No. layers in block for LV & 1, 2\\
 	No. blocks of layers for LV & 1, 2, 3, 4\\
 	Conv. filter size in LV & 2, 4, 6, 8, 16\\
 	MP layer kernel size for LV & 3, 5\\
 	MP layer stride length for LV & 1, 2, 3\\
 	LV input vector size & 151, 201, 251\\
 	No. FC layers & 1, 2, 3, 4\\
 	FC layer filter size & 64, 128, 256, 512, 1024\\
	\hline
	\hline
	\end{tabulary}
\end{table}

Since a CNN can only take a fixed-size input, two key parameters in the network architecture are the sizes of the input vector time series.
We adopted sizes of 2001 and 201 for the global and local views, respectively, which \citet{Shallue2018} found to be optimal for Kepler data. However, NGTS is a ground-based survey with a much shorter baseline and exposure time compared to Kepler. 
In order to see if a different vector size may improve performance we tested a full combination of 1001, 2001 and 3001 vector input sizes for the global view; and 151, 201 and 251 sized input vectors for the local view. 
Additional network parameters may also affect performance, so for each combination of view size we used the \textsc{Hyperopt} \citep{Bergstra2013} package, with Tree-structured Parzen Estimator (TPE) algorithm, to conduct a Bayesian optimisation over the model hyperparameter space.
We considered 19 hyperparameters (Table \ref{tab:optimisation_params}), including those associated with training (e.g. learning rate, dropout probability, number of epochs) and network architecture (e.g. kernel size, quantities of different layers).

Thousands of models were evaluated in total, spanning hundreds of hours of computation time, using NVIDIA Tesla P100 GPUs.
Each model took on average 11 minutes to train using all inputs, and 6 minutes using only global and local primary flux inputs.
Overall, we found no statistically significant improvement in performance of the network for alternative input vectors sizes or other hyperparameters. However, we note that we were only able to search an extremely small area of the overall hyperparameter space, due to resource limitations.
Further work is needed to clarify whether a different network architecture could boost performance for NGTS.

\subsection{Network Training}
\label{sub:NetworkTraining}

Finally, after completing the architecture search, we trained PlaNET on the different datasets we constructed using both real and simulated data.
We trained using a batch size of 50, a learning rate of $1x10^{-5}$ and for a maximum of 20 epochs. 
We employed early stopping to prevent over fitting, if the generalisation loss exceeded 20\%. We refer the reader to \citet{Prechelt2012} for a detailed discussion on early stopping. 
In short this meant that if the error on the validation set after any epoch exceeded the smallest error over all previous epochs by 20\% or more, training was immediately stopped. 
During training, the \textsc{Adam} optimisation algorithm \citep{Kingma2014} with default decay rates was utilised to minimise the cross-entropy loss function. 
To further prevent overfitting, dropout regularisation with a probability of 0.5 was applied to the fully connected layers, which acts to deactivate random neurons with some probability for the pass of every batch \citep{Hinton2012}.
We employed model averaging in the form of k-fold cross validation, to increase the reliability of our results. 
We achieved this by splitting every dataset into 10 segments, with 80\% of the segments used for training, 10\% for validation and 10\% for testing at any one time.
This corresponds to 24,000 lightcurves for training, 3,000 for validation and 3,000 for testing respectively. 
Ten different copies of each model were trained by rotating the segment used for validation and testing, while keeping the remaining ones for training. 
Additionally, a different random seed value was used each time.
The mean predictions from each of the 10 copies are then adopted as the final values. 

\section{Training with Simulated Data}
\label{sec:SimulatedData}

Using the procedure described in \S \ref{sub:NetworkTraining}, a neural network was trained on 100,000 fully-simulated NGTS lightcurves, generated as discussed in \S \ref{sub:SimulatedData}.
We consider four metrics for determining network performance:
\begin{itemize}
    \item \textbf{AUC:} Area under the receiver operating characteristic curve. This can be interpreted as the probability that a randomly chosen planet scores more highly than a randomly chosen false positive.
    \item \textbf{Accuracy:} The fraction of network classifications which are correct.
    \item \textbf{Precision:} The fraction of correctly classified planets over the total number of candidates classified as planets. 
    \item \textbf{Recall:} The fraction of planets which are recovered by the network.
\end{itemize}

The network achieved an AUC score of 98.82\%, an accuracy of 95.31\%, precision and recall of 99.18\% and 91.34\% respectively on the unseen test data. 
The high performance of the network on simulated data is encouraging, indicating that the neural network has the capacity to perform the classification task well.

\citet{Pont2006} have shown that correlated noise is complex and that this can significantly reduce the transit recovery rate. 
In order to quantify the effect of noise in the NGTS data, we compare two models: one trained using real data (with planetary transit and EB injections) and one trained using fully simulated data, under similar conditions. 
As explained in \S \ref{sec:DatasetPreparation}, the simulated data consists of pure noise lightcurves for the non-planet class, and noise plus injected transits for the transit-class.
 Therefore, for real data it most closely resembles the NP dataset (\S \ref{sub:RealData}) and so we use this as the basis of comparison between the two. 
To draw a valid comparison we use only 24,000 simulated data lightcurves for training, equal to the number of lightcurves in the real datasets. 
Training on more data is likely to increase performance, which we explore in more detail in (\S \ref{sub:DatasetSize}).

Re-training the neural network using only 24,000 simulated lightcurves, the model achieves an AUC of 98.12\% and an accuracy of 94.38\%. 
In contrast, the NP dataset achieves an AUC of 96.00\% and an accuracy of 90.10\% respectively.

Reduced performance when training on real data appears to support our hypothesis that the systematic noise properties of the real data are more complex than that modelled for our simulated data. 

In order to draw a more direct comparison, we further investigated how well a network trained on simulated data performs when classifying real data. 
We trained a model using 100,000 simulated lightcurves and subsequently classified the NP test dataset. The result was an AUC of 85.0\% and an accuracy of 80.1\%, measured over 2,000 lightcurves. In this case, performance is worse than when the models are trained and validated on the same dataset compositions.

These results highlight the main issue with training a neural network using simulated data. Previous works \citep{Shallue2018,Dattilo2019} have made efforts to remove data artefacts and systematic effects prior to passing the data through the network. The assumption being that this boosts performance. However, \citet{Zucker2018} noted that CNNs are theoretically capable of learning the noise properties of the data. Future work may reveal the extent to which this is true.

\subsection{Dataset Size} 
\label{sub:DatasetSize}

\begin{figure}
	\centering
	\includegraphics[width=0.95\columnwidth]{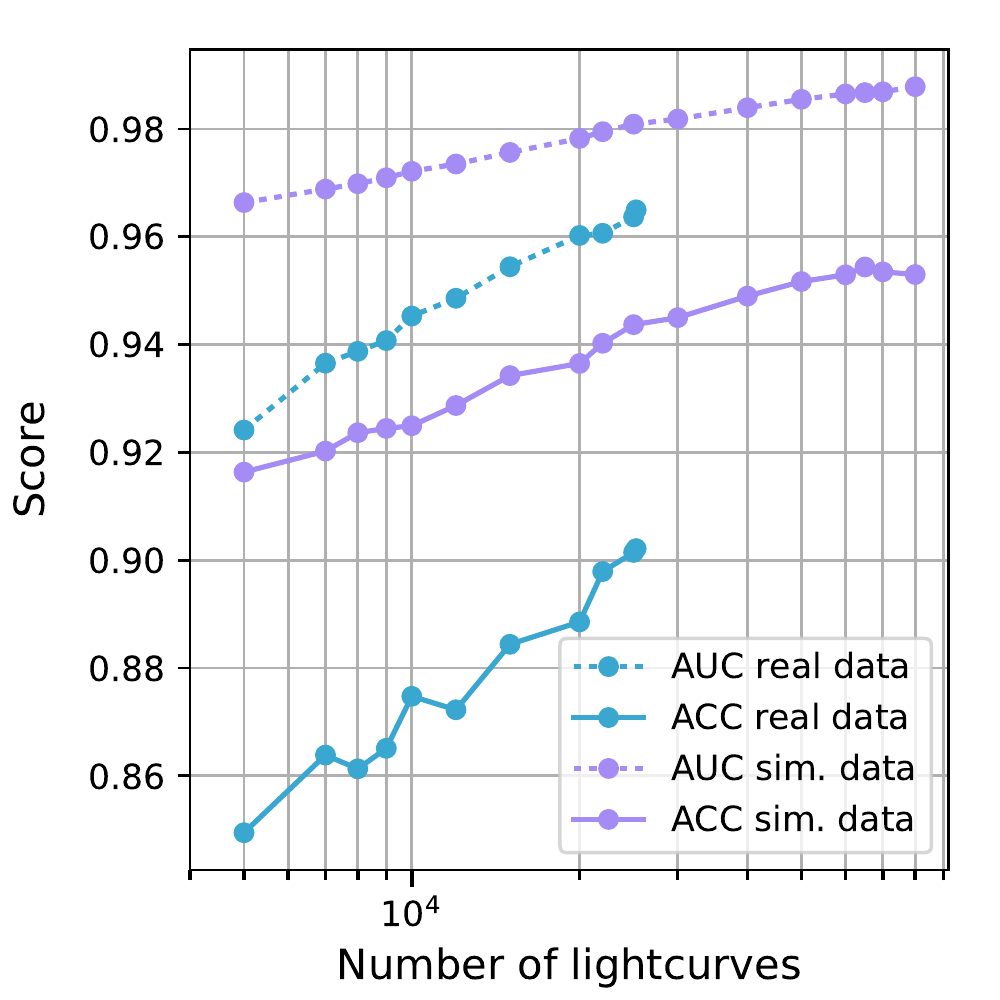}
    \caption{Network AUC (dashed line) and accuracy (solid line) metrics as a function of the training dataset size, for fully simulated (teal data points) and real NGTS data with planetary transit and EB injections (purple data points). Datasets contain only the non-periodic subclass of lightcurves in the non-planet class. Performance for simulated data is sampled at larger dataset sizes due to its increased ease of production. Quantities were measured over the 10\% test dataset, which was not used during training. The learning rate and number of training epochs were fixed at $1\times10^{-5}$ and 20 respectively. For each metric, the network trained on simulated data scores more highly than training on real NGTS lightcurves (NP dataset), irrespective of dataset size. AUC and accuracy are positively correlated with the size of the training dataset, although the gradient for real data is steeper. The higher initial performance of the simulated data requires that any performance increase has to be made for low S/N transits. This likely explains the difference in gradient, as the distribution of S/N is the same for all dataset sizes. 
    }
    \label{fig:SimTrainingSize}
\end{figure}

Given the large volume of simulated data available, we investigated network performance as a function of the training dataset size. 
The results can be seen in Fig. \ref{fig:SimTrainingSize}, compared with the NP dataset for up to 24,000 lightcurves. 
Performance, as measured by both AUC and accuracy metrics, clearly increases when training on more lightcurves. 
Curiously, the performance increases faster for the NP dataset compared to the simulated data.

The higher initial performance for the network trained on simulated data means that any gains made must be in the low signal to noise regime, which may explain why the neural network improves more slowly.
An example of the network performance as a function of S/N can be seen in \S \ref{sec:RealResults}.
As expected, most of the misclassifications are for very shallow transits which are harder to correctly identify.

A side effect of this behaviour is that the performance metrics of the neural network are correlated with the distribution of transit signal to noise, though not in a trivial way. 
For example, increasing the number of shallow transits with S/N < 5 may lower the performance as the network will struggle to recover them, but this will be somewhat compensated for by the improved performance from the larger training set size. 
This points to the difficulty of comparing the performance of different neural networks using the AUC and other metrics alone, without fixing the underlying distributions of the data. 

Finally, Fig. \ref{fig:SimTrainingSize} shows that additional gains may be made by increasing the dataset size beyond what is currently being used. 
Simulated data is useful as the dataset size is only constrained by how much time is spent producing each lightcurve, so one potential strategy may be to use `transfer learning' whereby the neural network is trained on simulated data first and then subsequently trained with real data. 
This was tried, however the performance improvement was very small.

\subsection{Label Noise} 
\label{sub:label_noise}

\begin{figure}
	\centering
	\includegraphics[width=0.95\columnwidth]{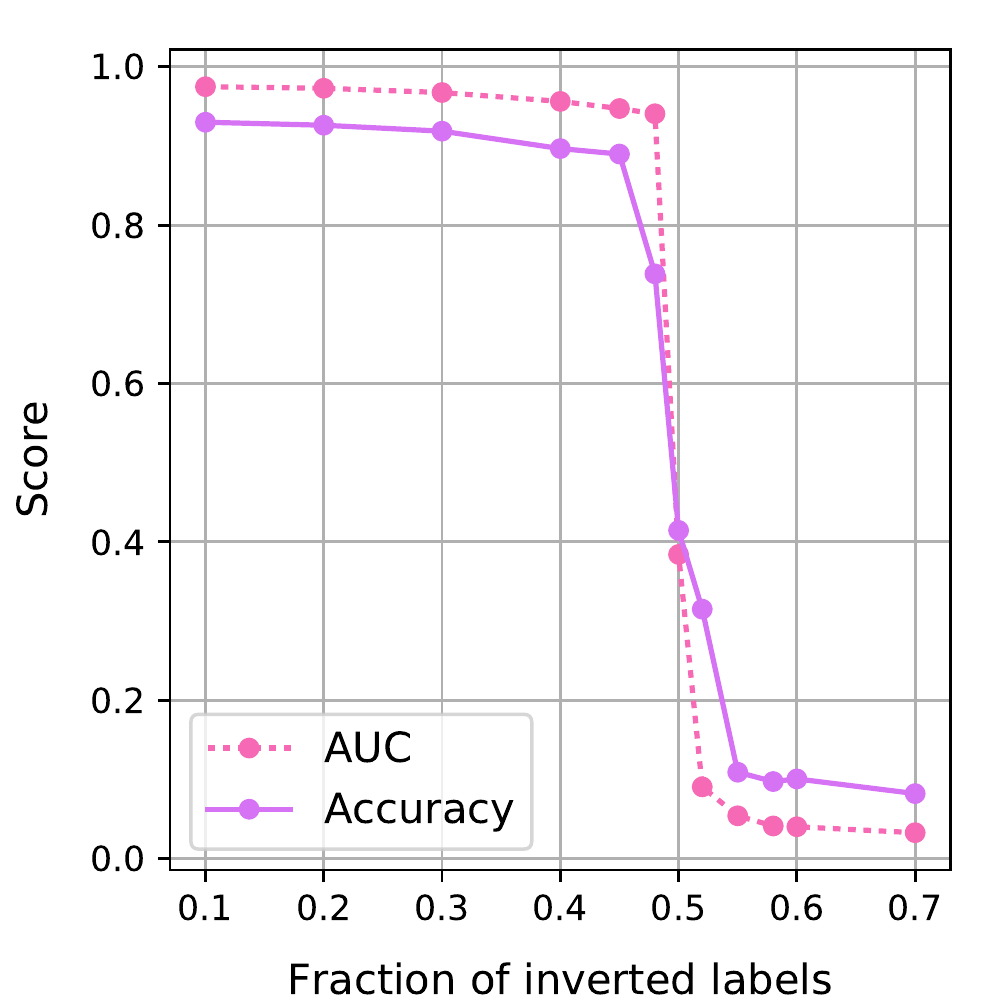}
    \caption{Network AUC (pink data points with dashed lines) and accuracy (purple data points with solid lines) metrics for the fully simulated dataset comprising 24,000 lightcurves, as a function of the fraction of deliberately mislabelled lightcurves in the training dataset.
    Quantities were measured over the 10\% test dataset, whose labels are unchanged. The learning rate and number of training epochs were fixed at $1\times10^{-5}$ and 20 respectively. For both metrics, there is minimal impact on performance up to a inverted label fraction of around 0.45, with a steep decline after. Above 0.5 there is a label inversion and the performance of the network approaches zero (within errors) on the test set.}
    \label{fig:SimLabelNoise}
\end{figure}

As we discussed in \S \ref{sec:DatasetPreparation}, one potential issue with using real lightcurves for training is that there may be contamination from real low S/N transit events. 
That is to say, lightcurves may have incorrect class labels.
Fully-simulated data provides a pristine environment in which to test the effect of this contamination, as the ground truth is definitively known for each case. 

Using the simulated dataset, we explored our network's susceptibility to `noise' in the training dataset class labels. 
We achieved this by inverting a varying percentage of labels prior to passing the lightcurves through the network i.e. a proportion of class labels were changed from 0 to 1 and vice-versa. 
The performance of the network was then measured on the test set, which had not been altered in any way. 
Results are presented in Fig. \ref{fig:SimLabelNoise}.

It can be seen that performance degrades linearly up to a contamination fraction of approximately 45\%, after which it declines rapidly.
The loss in accuracy up to 45\% contamination was $\sim$4\%. 
\citet{Reis2019} perform the same experiment for probabilistic random forests and found a loss of less than 5\% when more than 45\% of their dataset had incorrect labels, in-line with the performance drop we find.
Evidently label contamination does hinder performance, but the network is robust to small contamination fractions. 
Levels of label contamination for the real datasets are likely to be low, thus network performance when training on real data is not significantly impacted. 
Our findings are consistent with results from other work showing that CNNs are robust to label noise \citep{Rolnick2017,Li2019}.

We also conclude that label contamination is unlikely to be a major contributing factor as to why our network trained on simulated data, achieved better performance compared to training on real data, which we discussed in \S \ref{sec:SimulatedData}.

\section{Training with NGTS Data} 
\label{sec:RealResults} 

As we have shown in \S\ref{sec:SimulatedData}, training on simulated data alone is not sufficient to achieve the best possible performance of the neural network.
In this section we expand on results obtained when training PlaNET using real NGTS lightcurves.
Table \ref{tab:Dataset_perf} shows the AUC, accuracy, precision and recall for each dataset composition, trained using the procedure outlined in \S \ref{sub:NetworkTraining} and measured on test datasets. 
The OFP model performs best in training with an AUC and accuracy of $99.3 \pm{0.2}$ \% and $95.8 \pm{0.5}$ respectively, compared with the remaining five models which likewise score approximately $96.0$\% and $90.0$\% respectively. 
These scores are broadly consistent with other studies \citep{Shallue2018,Dattilo2019}. 
Models which contain \textsc{ORION} false positives have many high S/N candidates in the non-planet class, as these are preferentially selected by \textsc{ORION}.
This may account for why models containing \textsc{ORION} false positives score more highly.

For the \orthree\ and \ortwo\ models, the datasets using the Max and Random selection criteria perform equally well, while for the \orone\ case the Min variant is best.
However, the differences between the models are relatively small and within errors. 
It can be seen from Table \ref{tab:Dataset_perf} that the best overall model for classifying NGTS lightcurves is OFP Min.

\begin{figure*}
	\centering
	\includegraphics[width=0.95\textwidth]{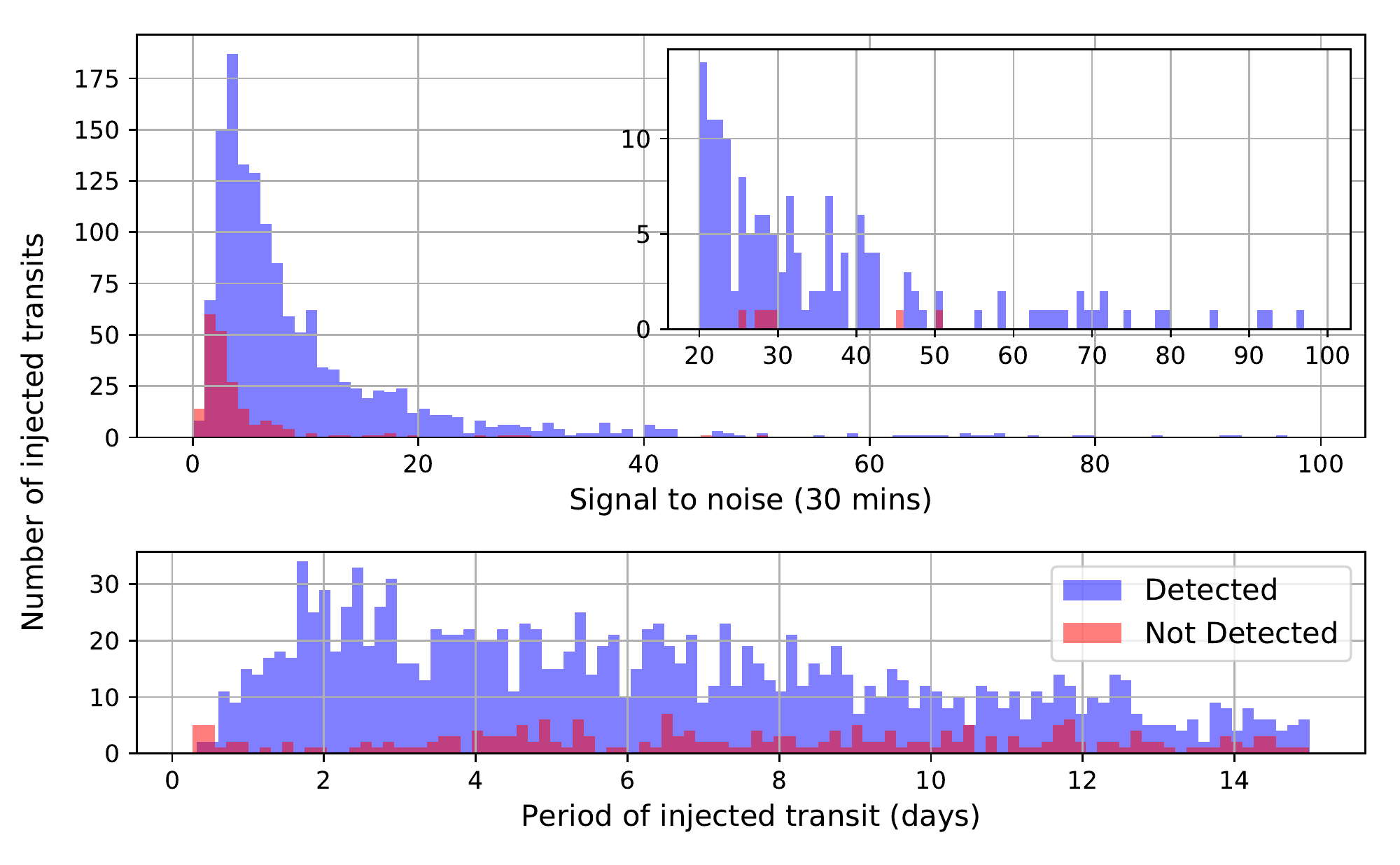}
    \caption{\textit{Top panel:} Histogram of detected and non-detected transits by the network, as a function of S/N. Performance is measured on the 10\% test component of the real NGTS \ortwo\ Max model. The S/N is calculated as the transit depth divide by the standard deviation before transit injection, after phase folding the time-series to the correct period and binning in exposure time to 30 minutes. The inset figure shows a zoomed in view of the high S/N value range. The distribution of injected transits is biased towards low S/N values, because there are far more faint host lightcurves, which are comparatively noisy. As expected the vast majority of undetected transits have low S/N values, however the network also fails to detect a small number of transits with high S/N.
    \textit{Bottom panel:} Similar to the top panel, but for the period of injected transits as opposed to the signal to noise. The distribution of periods is slightly skewed towards shorter values, where it is more likely that a trial transit injection will meet our validation criteria of having at least three transits, each covering at least one third of a transit. The fraction of undetected transits is higher for longer periods. This is because phase folding increases the S/N of the transit signal, but at larger periods there are fewer individual transits available, so the benefits of phase folding are diminished.
}
    \label{fig:DetectFrac}
\end{figure*}

Fig. \ref{fig:DetectFrac} shows the fraction of recovered transits as a function of signal to noise and period for one ensemble of the \ortwo\ Max model. 
Below signal to noise values of 10, the fraction of correctly classified transit lightcurves decreases progressively. 
This is expected behaviour as lower signal to noise transits will be harder to distinguish from noise. 
It is particularly obvious for S/N lower than 5, where the detection fraction reduces to 76.5\% compared to 95.4\% for higher values. 
Most of the decrease seen below S/N of 5 is due to transits with a S/N value less than 2.0, where the detection fraction is 50.3\%, while in the 2-5 S/N range the network still manages a detection fraction of 83.5\%. 
As can be seen in the inset of Fig. \ref{fig:DetectFrac} there are several deep transits which are incorrectly classified.
These transits should be easy to identify, even prior to phase folding the lightcurve, however they are misclassified by the network. 

Table \ref{tab:SDE} gives a breakdown of the performance of the different models containing \textsc{ORION} false positives as a function of the S/N of the injected transits. 
These are calculated as the mean across all 10 ensembles.
The \ortwo\ and \orthree\ models have the highest number of non-recovered high S/N transits, while the \orone\ model performs best.
The \orone\ model does not contain any non-periodic lightcurves, which may be hard to distinguish from lightcurves injected with shallow transits.
Furthermore, the large number of \textsc{ORION} false positives in the \orone\ dataset may make it easier to separate the transits in general. 
There are no statistically significant variations in the number of false negatives within the different SDE variations of each dataset.
For the \ortwo\ and \orone\ models the number of transits not recovered at high S/N is greater than that in the medium S/N range. 
This is paradoxical as we would expect the former to be easier to detect than the latter.
No obvious features were present in these high S/N transits which might explain why the were not correctly classified. 
Our current hypothesis is that it is necessary to increase the number of examples of such transits in the training data. 
In practise this is limited by the number of bright stars in the dataset, as we do not want to inject physically unrealistic planets.

\begin{table*}
	\centering
	\caption{Network performance when training on the different real NGTS datasets, which differ in the compositions of the non-planet class. Performance is measured on the their respective 10\% unseen test dataset components of similar composition.
	\label{tab:Dataset_perf}
	Accuracy, precision and recall are based on a probability threshold of 0.5; AUC is independent of threshold. For models containing the OFP category, there are four different versions corresponding to the different SDE selection methods of \textsc{ORION} false positive candidates. Uncertainties are derived from k-fold cross validation, using 10 model training repetitions with a different random seed and portion of the dataset. The model performing best in training is highlighted in bold.}
	\begin{tabulary}{0.90\linewidth}{LLCCCC}
	{Model} & {OFP selection} & {AUC} & {Accuracy} & {Precision}  & {Recall}\\
	\hline
    \orone & Max & $0.992\pm{0.002}$ & $0.956\pm{0.006}$ & $0.960\pm{0.011}$ & $0.960\pm{0.011}$\\
    & \textbf{Min} & $\mathbf{0.994\pm{0.000}}$ & $\mathbf{0.964\pm{0.002}}$ & $\mathbf{0.974\pm{0.003}}$ & $\mathbf{0.974\pm{0.003}}$\\
    & Uniform & $0.993\pm{0.001}$ & $0.960\pm{0.002}$ & $0.968\pm{0.004}$ & $0.968\pm{0.004}$\\
    & Random & $0.993\pm{0.000}$ & $0.960\pm{0.002}$ & $0.967\pm{0.004}$ & $0.967\pm{0.004}$\\
    \ortwo & Max & $0.958\pm{0.002}$ & $0.886\pm{0.002}$ & $0.902\pm{0.006}$ & $0.902\pm{0.006}$\\
    & Min & $0.954\pm{0.002}$ & $0.882\pm{0.002}$ & $0.906\pm{0.007}$ & $0.906\pm{0.007}$\\
    & Uniform & $0.955\pm{0.001}$ & $0.883\pm{0.002}$ & $0.905\pm{0.006}$ & $0.905\pm{0.006}$\\
    & Random & $0.958\pm{0.001}$ & $0.887\pm{0.002}$ & $0.907\pm{0.005}$ & $0.907\pm{0.005}$\\
    \orthree & Max & $0.956\pm{0.002}$ & $0.885\pm{0.003}$ & $0.900\pm{0.006}$ & $0.900\pm{0.006}$\\
    & Min & $0.953\pm{0.001}$ & $0.881\pm{0.002}$ & $0.904\pm{0.006}$ & $0.904\pm{0.006}$\\
    & Uniform & $0.954\pm{0.002}$ & $0.882\pm{0.002}$ & $0.905\pm{0.006}$ & $0.905\pm{0.006}$\\
    & Random & $0.957\pm{0.001}$ & $0.886\pm{0.002}$ & $0.903\pm{0.005}$ & $0.903\pm{0.005}$\\
    \nwnextgen &   & $0.968\pm{0.001}$ & $0.903\pm{0.001}$ & $0.924\pm{0.004}$ & $0.924\pm{0.004}$\\
    \nextgen &   & $0.958\pm{0.002}$ & $0.891\pm{0.002}$ & $0.908\pm{0.004}$ & $0.908\pm{0.004}$\\
    \blank &   & $0.960\pm{0.001}$ & $0.901\pm{0.002}$ & $0.933\pm{0.005}$ & $0.933\pm{0.005}$\\
	\hline
	\hline
	\end{tabulary}
\end{table*}

\begin{table*}
	\centering
	\caption{Percentage of false negatives for models trained using the 12 real NGTS datasets containing OFPs, as a function of the injected transit S/N, evaluated over their 10\% test dataset components. The mean and standard error values are calculated over the ensemble of 10 models trained for each dataset as discussed in \S \ref{sub:NetworkTraining}. The S/N values are taken as the transit depth divided by the standard deviation of the phase-folded lightcurve binned to 30 minute cadence. The standard deviation is calculated prior to injection of the transit. The different SDE variations of each model have false negative fractions within the statistical errors of each other. However, the \orone\ model performs better than the \ortwo\ and \orthree\ models in the high and low S/N regimes. On the low S/N end this may be because there are no non-periodic lightcurves included, which are difficult to distinguish from shallow transits. Furthermore, the inclusion of a large number of \textsc{ORION} false positives may make it easier to distinguish between transits and non-transits in general. This perhaps explains the better performance in the high S/N regime as well.}
	\label{tab:SDE}
	\begin{tabulary}{0.90\columnwidth}{llrrr}
	{Model} & {SDE} & {S/N > 20} & {10 < S/N < 20} & {S/N < 10}\\
	\hline
	\orone & Max & $4.2 \pm{0.8}$ &  $3.3 \pm{0.6}$ & $5.2 \pm{0.4}$ \\
	 & Uniform & $3.1 \pm{0.5}$ &  $2.5 \pm{0.5}$ & $5.1 \pm{0.3}$ \\
	 & Random & $3.6 \pm{0.4}$ &  $3.1 \pm{0.5}$ & $5.5 \pm{0.5}$ \\
	 & Min & $2.8 \pm{0.5}$ &  $2.7 \pm{0.4}$ & $5.4 \pm{0.3}$ \\
	 NP/EB/OFP/WF & Max & $7.2 \pm{1.0}$ & $3.3 \pm{0.3}$ & $17.0 \pm{0.8}$ \\ 
	 & Min & $7.2 \pm{0.8}$ & $3.6 \pm{0.4}$ & $19.3 \pm{0.8}$ \\
	 & Uniform & $6.5 \pm{1.0}$ & $3.1 \pm{0.4}$ & $17.5 \pm{0.9}$ \\
	 & Random & $7.2 \pm{0.9}$ & $3.5 \pm{0.4}$ & $19.2 \pm{1.0}$ \\
	NP/EB/OFP & Max & $6.7 \pm{0.9}$ & $3.2 \pm{0.3}$ & $16.6 \pm{0.8}$ \\ 
	 & Min & $7.5 \pm{1.1}$ &  $4.0 \pm{0.5}$ & $19.1 \pm{0.8}$ \\
	 & Uniform & $6.6 \pm{0.9}$ & $3.2 \pm{0.3}$ & $16.8 \pm{0.8}$ \\
	 & Random & $7.0 \pm{1.0}$ & $3.6 \pm{0.5}$ & $18.4 \pm{0.9}$ \\
	\hline
	\hline
	\end{tabulary}
\end{table*}

\section{Comparison to NGTS Eyeballing}
\label{sec:EyeComparison} 

\begin{table*}
	\centering
	\caption{Model performance when training on real NGTS data, measured as per Table \ref{tab:Dataset_perf} but compared to lightcurve flags assigned during the vetting process. Correct predictions from the network constitute a probability greater than 0.5 for flags P, AD, AS and BS, and less than 0.5 for the remaining flags. The low precision of the models is due to the unbalanced nature of the problem, with planets and manually selected promising candidates only making up $\sim$1\% of the candidates. Therefore even with a relatively low false positives rate, the number of false positives would greatly outnumber the true candidates resulting in a very low precision.}
	\label{tab:DatasetPerf}
	\begin{tabulary}{0.90\linewidth}{LLCCCC}
	{Model} & {OFP selection} & {AUC} & {Accuracy} & {Precision}  & {Recall}\\
	\hline
	\orone & Max & $0.779\pm{0.004}$ & $0.877\pm{0.009}$ & $0.0137\pm{0.0004}$ & $0.42\pm{0.02}$\\
	 & Min & $0.737\pm{0.005}$ & $0.894\pm{0.007}$ & $0.0132\pm{0.0007}$ & $0.341\pm{0.007}$\\
	 & Uniform & $0.770\pm{0.004}$ & $0.902\pm{0.008}$ & $0.0147\pm{0.0007}$ & $0.35\pm{0.02}$\\
	 & Random & $0.765\pm{0.005}$ & $0.906\pm{0.009}$ & $0.0144\pm{0.0006}$ & $0.33\pm{0.02}$\\
	
	\orthree  & Max & $0.775\pm{0.005}$ & $0.776\pm{0.009}$ & $0.0106\pm{0.0003}$ & $0.60\pm{0.02}$\\
	  & Min & $0.715\pm{0.005}$ & $0.804\pm{0.015}$ & $0.0094\pm{0.0004}$ & $0.45\pm{0.02}$\\
	 & Uniform & $0.764\pm{0.004}$ & $0.797\pm{0.009}$ & $0.0109\pm{0.0003}$ & $0.56\pm{0.01}$\\	
	  & Random & $0.748\pm{0.004}$ & $0.836\pm{0.010}$ & $0.0112\pm{0.0004}$ & $0.46\pm{0.02}$\\
	
	\ortwo & Max & $0.765\pm{0.004}$ & $0.746\pm{0.011}$ & $0.0098\pm{0.0002}$ & $0.63\pm{0.02}$\\
	 & Min & $0.721\pm{0.005}$ & $0.753\pm{0.015}$ & $0.0084\pm{0.0003}$ & $0.52\pm{0.02}$\\
	 & Uniform & $0.761\pm{0.003}$ & $0.766\pm{0.010}$ & $0.0101\pm{0.0003}$ & $0.60\pm{0.02}$\\
	 & Random & $0.746\pm{0.006}$ & $0.799\pm{0.011}$ & $0.0102\pm{0.0002}$ & $0.52\pm{0.02}$\\

	\nextgen & & $0.652\pm{0.004}$ & $0.417\pm{0.011}$ & $0.0054\pm{0.0001}$ & $0.81\pm{0.02}$\\
	\nwnextgen & & $0.639\pm{0.004}$ & $0.382\pm{0.011}$ & $0.0053\pm{0.0001}$ & $0.84\pm{0.01}$\\
	\blank & & $0.503\pm{0.006}$ & $0.094\pm{0.005}$ & $0.0039\pm{0.0001}$ & $0.913\pm{0.009}$\\
	\hline
	\hline
	\end{tabulary}
\end{table*}

As discussed in \S \ref{sub:RealData} the NGTS dataset used in this paper consists of 91 fields, 890,000+ lightcurves and detections of a transit event in 58,500+ targets.
At the time of writing two fields have not yet been vetted, these were excluded from our analysis.
For the remaining fields, 3,042 detections were classified as either a promising candidate or clear false positive.
This presents an opportunity to compare the performance of the neural network classifications in detail to that of expert human vetters.

For each of these targets, \textsc{ORION} produces up to 5 separate detections at different periods and epochs, corresponding to the top-5 peaks in the box least-squares periodogram. 
Each peak corresponds to a candidate which we classify using PlaNET, trained with all of the datasets in \S \ref{sec:RealResults}, summarised in Table \ref{tab:dataset_summary}.
For completeness we included candidates with periods greater than 15.0 days in our performance evaluation, despite not including these in the training data.
We remind the reader that for dataset compositions which include ORION false positives in the non-planet class, we divided the data into two groups based on their NGTS field. 
We created 2 versions of each dataset, drawing OFPs from the respective groups.
This is to ensure that PlaNET has not been trained on the same lightcurves it is later evaluating. 
For each classification we take the mean of the probability coming from each of the ten different copies of the model (trained with a different random seed and a different data fold).

\subsection{Eyeballing flags}
\label{sub:eyeballing_flags}

\begin{table*}
	\centering
	\caption{Fraction of correct classifications of \textsc{ORION} candidates by the neural network, as a function of lightcurve flag assigned during the vetting process. Lightcurves with flags AD, AS, BS and D are considered correctly classified if the network predicts probabilities greater than 0.5. For the remaining flags, a correct classification requires probabilities less than or equal to 0.5. Uncertainties are derived from k-fold cross validation, using 10 model training repetitions with a different random seed and portion of the dataset. Results are presented for models trained on different real NGTS datasets. For models with \textsc{ORION} false positives, we present results from the Max SDE variant. We determine that the best model, giving optimal balance between recovery of transits and a low false positive rate, is the \ortwo\ model, highlighted in bold. The motivation for choosing this model was to ensure as many of the AD, AS and BS candidates are recovered as possible. In practise, minimising the risk of missing a promising candidate is more important than reducing the false positives by a few additional percent.}
	\label{tab:FlagPerformance}
	\begin{tabulary}{\linewidth}{lRRRRRRRRRR}
	{Model} & {AD} & {AS} & {BS} & {D} & {EA1} & {EA2} & {EB} & {OTH} & {SINE} & {No Flag}\\
	\hline
	\orone & $0.627\pm{0.027}$ & $0.321\pm{0.021}$ & $0.332\pm{0.026}$ & $0.302\pm{0.016}$ & $0.671\pm{0.028}$ & $0.796\pm{0.029}$ & $0.942\pm{0.010}$ & $0.870\pm{0.009}$ & $0.920\pm{0.007}$ & $0.877\pm{0.009}$   \\
	\orthree & $0.825\pm{0.011}$ & $0.485\pm{0.022}$ & $0.489\pm{0.023}$ & $0.515\pm{0.015}$ & $0.692\pm{0.014}$ & $0.848\pm{0.008}$ & $0.937\pm{0.003}$ & $0.771\pm{0.009}$ & $0.909\pm{0.006}$ & $0.767\pm{0.009}$   \\
    \rowstyle{\bfseries\boldmath}
    \ortwo & $0.855\pm{0.014}$ & $0.544\pm{0.025}$ & $0.521\pm{0.027}$ & $0.566\pm{0.018}$ & $0.677\pm{0.013}$ & $0.839\pm{0.008}$ & $0.949\pm{0.002}$ & $0.730\pm{0.011}$ & $0.926\pm{0.008}$ & $0.744\pm{0.011}$ \\
	\nextgen & $0.968\pm{0.003}$ & $0.726\pm{0.027}$ & $0.737\pm{0.026}$ & $0.805\pm{0.010}$ & $0.243\pm{0.009}$ & $0.298\pm{0.015}$ & $0.415\pm{0.007}$ & $0.229\pm{0.002}$ & $0.338\pm{0.008}$ & $0.413\pm{0.011}$ \\
	\nwnextgen & $0.971\pm{0.002}$ & $0.774\pm{0.018}$ & $0.775\pm{0.020}$ & $0.836\pm{0.010}$ & $0.219\pm{0.009}$ & $0.282\pm{0.016}$ & $0.374\pm{0.008}$ & $0.214\pm{0.006}$ & $0.292\pm{0.008}$ & $0.378\pm{0.011}$ \\
	\blank & $0.996\pm{0.002}$ & $0.892\pm{0.015}$ & $0.861\pm{0.012}$ & $0.959\pm{0.005}$ & $0.006\pm{0.000}$ & $0.005\pm{0.000}$ & $0.021\pm{0.002}$ & $0.042\pm{0.005}$ & $0.100\pm{0.007}$ & $0.090\pm{0.005}$  \\
	\hline
	\hline
	\end{tabulary}
\end{table*}

Table \ref{tab:DatasetPerf} shows the level of agreement between model predictions and flags assigned by expert vetters. 
We define the agreement for positive class flags (P, AS, BS, AD, D) as those receiving network probabilities greater than 0.5, or those receiving 0.5 or less in the case of false positive flags (EA1, EA2, EB, OTH, SINE, UNF, No flag). 
Candidates which have been unflagged are included among the negative labels. 
This is conservative, as being unflagged means that at least one human eyeballer thought the candidate was interesting enough to be discussed, but other eyeballers were not convinced by its legitimacy.
Targets without flags are also considered to be part of the negative class, as the vast majority are expected to be false positives from yield studies \citep{Gunther2017a} and from ongoing follow-up work.

From Table \ref{tab:DatasetPerf} we see that models with no \textsc{ORION} false positive subclass to their training dataset, perform poorly compared to those which include them.
This is in contrast to performance measured on the unseen test dataset, which showed relatively similar AUC values.
This is not surprising since models containing \textsc{ORION} false positives (\orone, \ortwo\ and \orthree\ models) more closely resemble the candidate lightcurves which have been evaluated.

However, unlike in Table \ref{tab:Dataset_perf}, the performance of the \orone\ model is not better than the \ortwo\ or \orthree\ models. 
Instead they achieve a very similar performance, despite the \ortwo\ and \orthree\ models containing fewer false positives. 
It is also worth noting that the Max version of each model perform best across all three datasets. 

The precision of models measured using eyeballing labels is not as informative as when evaluated on the test dataset. 
For the former, the precision is at best only 1\%, but this is of little concern. 
Precision means the fraction of candidates with probabilities greater than 0.5, which also have one of the following flags: `P', `AS', `BS', `AD', `D'. 
The sample of candidates with such flags constitute only 1\% of the total population, but we showed in \S\ref{sec:RealResults} that the false positive rate is 10\%. Therefore even in the best case scenario, where every true positive found by PlaNET had been flagged as a promising candidate, the precision would still only be 9\%.
Put another way, there are many more false positive \textsc{ORION} candidates in the dataset than promising candidates, so even a low false positive rate would reduce the precision substantially.

Table \ref{tab:FlagPerformance} also shows the agreement between the flag assigned by NGTS vetters and the neural network, this time for specific flags and for six out of the fifteen models. 
Models with a larger proportion of false positives perform worse in selecting AD, AS, BS or D candidates correctly. 
Conversely the models with no false positives perform much worse when correctly identifying the various false positive labels and the candidates with no given flag.
The proportion of false positives included appears to bias the network towards either being `strict' or `lenient' with regards to vetting the candidates.

Within the different models we note that overall performance is better for AD candidates than for AS or BS candidates. AD candidates have deeper transits and so have a higher S/N than AS or BS candidates and are therefore easier to classify. This is consistent with the results in Fig. \ref{fig:InjectedDistributions} which show that the detection fraction decreases as the S/N value decreases.

\subsection{Confirmed planets}
\label{sub:confirmed_planets}

At the time of writing, the NGTS dataset contains lightcurves for 14 confirmed planets, with 10 of those discovered by NGTS and 4 other planets which happened to fall within the NGTS fields. 
Table \ref{tab:planet_probs} shows the network probability values for each of these planets. 
The Max dataset versions have been adopted for models which contain \textsc{ORION} false positives in the non-planet class, as Table \ref{tab:DatasetPerf} shows this model performs better than the three alternatives.
From left to right, models in Table \ref{tab:planet_probs} contain an increasing number of false positives, which is correlated with a decrease in the number of recovered planets. 
Taking NGTS-2b \citep{Raynard2018} as an example, the network predicts a lower planetary probability as more false positives are included in the non-planetary class. 
This effect culminates with the OFP model, comprised entirely of false positives in the negative class, failing to recover additional planets. 
We could not discern an obvious reason as to why the network struggles to recover NGTS-2b in particular. 
With a 1\% transit depth, this planet should be easily identifiable in the light curve. 
In fact, the precision of the NGTS lightcurve is so high for this planet that it was confirmed from 9 individual transits without the need for follow-up photometry. Likewise, there was no obvious pattern to the planets not recovered by the OFP model. 

Conversely, models with no false positives in their training datasets perform best, recovering all of the known planets, even NGTS-4b \citep{West2018} which with a transit depth of $1.3\pm 0.2$~mmag, represents the shallowest detection of a transiting exoplanet from the ground with a wide-field survey. 
While this might imply that these models are overall superior, we note that their precision is much lower than models which include false positives. 
This adequately highlights the trade-off between reducing the false positive rate versus maximising the planet recovery rate.
Finally, we note that only probabilities from the Max dataset variants were shown. 
The Min, Random and Uniform variants consistently missed more confirmed planets than Max, with Uniform performing the worst.
There appears to be no consistent pattern in which planets were missed across the different SDE varients, making it difficult to explain why they were not recovered.

\begin{table*}
	\centering
	\caption{Predicted network probabilities for confirmed planets with NGTS lightcurves. Results are presented for models trained on different real NGTS datasets, which differ in their composition of the non-planet class.
	Planets with all-numerical designations are confirmed within the NGTS consortium but have not yet been published. Probabilities are the mean values averaged over 10 independent models, each trained with different portions of the overall dataset and different random seeds. For models containing false positives, we present results from the Max SDE variant. Models with no false positives are more optimistic, predicting high probabilities for all planets. In contrast, the other models predict probabilities below 0.5 for some planets, these cases are highlighted in bold.}
	\label{tab:planet_probs}
	\begin{tabulary}{0.9\linewidth}{LRRRRRR}
	{Planet Name} & {\blank} & {\nwnextgen} & {\nextgen} & {\ortwo} & {\orthree} & {\orone} \\
	\hline
    NGTS-1b \citep{Bayliss2018} & 0.993 & 0.996 & 0.992 & 0.992  & 0.991 & 0.986 \\
    NGTS-2b \citep{Raynard2018} & 1.000 & 0.970 & 0.970 & \textbf{0.122}  & \textbf{0.065} & \textbf{0.049} \\
    NGTS-3Ab \citep{Gunther2018} & 0.998 & 0.995 & 0.995 & 0.933 & 0.927 &  0.835 \\
    NGTS-4b \citep{West2018} & 0.981 & 0.981 & 0.981 & 0.771  & 0.709 & \textbf{0.391} \\
    NGTS-5b \citep{Eigmuller19} & 0.997 & 0.996 & 0.996 & 0.988  & 0.991 & 0.967 \\
    NGTS-6b \citep{Vines19} &  0.949 & 0.915 & 0.915 & 0.923 & 0.921 &  0.969 \\
    NOI-101123 (in prep) & 0.992 & 0.983 & 0.983  & 0.792 & 0.729 & 0.761 \\
    NOI-101155 (in prep) & 0.996 & 0.993 & 0.993  & 0.860 & 0.845 & \textbf{0.146} \\
    NOI-102329 (in prep) & 0.995 & 0.991 & 0.991  & 0.741 & 0.631 & \textbf{0.441} \\
    NOI-101635 (in prep) & 0.998 & 0.996 & 0.993  & 0.945 & 0.943 & 0.603 \\
    WASP-68b \citep{delrez14} & 1.000 & 0.999 & 0.999 & 0.676 & 0.524 & \textbf{0.042} \\
    WASP-98b \citep{Hellier14} & 0.992 & 0.992 & 0.992 & 0.935 & 0.888 &  0.94 \\
    WASP-131b \citep{Hellier17} & 0.972 & 0.783 & 0.783 & 0.782 & 0.780 & 0.864 \\
    HATS-43b \citep{Boisse13} & 0.999 & 0.998 & 0.994 & 0.786 & 0.685 & \textbf{0.273} \\
    \hline
    \hline
	\end{tabulary}
\end{table*}

\subsection{Probability distribution and thresholds}
\label{sub:prob_dist_and_thresh}

\begin{figure*}
	\centering
	\includegraphics[width=0.95\textwidth]{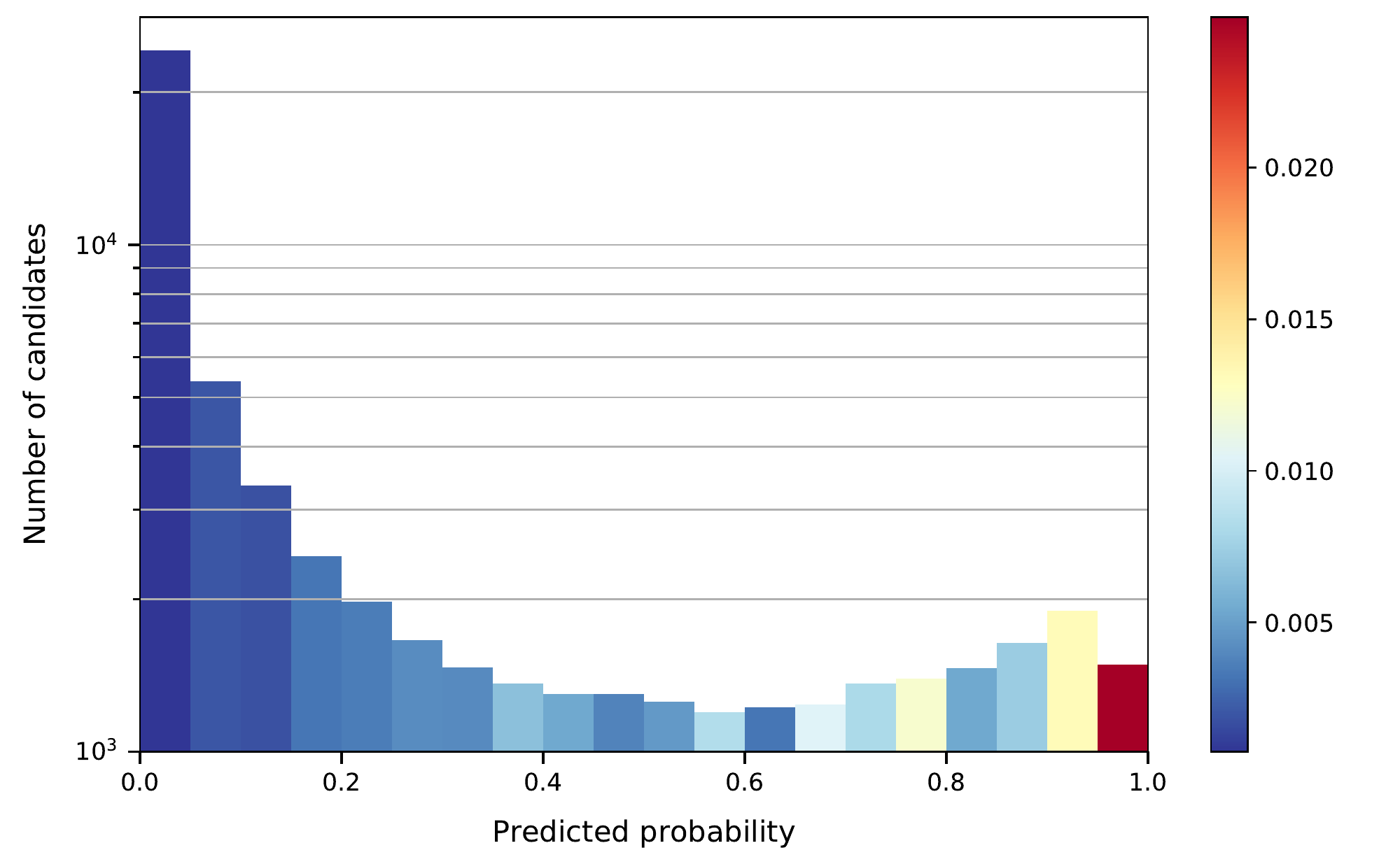}
    \caption{Histogram of probability predictions for \textsc{ORION} candidates and confirmed planets using the \ortwo\ Max model. Probabilities are the mean values averaged over 10 independent models, each trained with different portions of the overall dataset and different random seeds. The colour bar indicates the fraction of candidates in each bin which have AD, AS, BS or P flags. The majority of candidates receive either very high or very low probabilities, demonstrating that the network has good discriminatory power. There are a small number of candidates with probabilities close to the 0.5 threshold, for which the network is less certain. Over 50\% of the \textsc{ORION} candidates are given a probability of less than 0.1, which could be de-prioritised during the human vetting stage. Bins in the 0.9 to 1.0 range contain a larger fraction of promising candidates and confirmed planets, indicating good agreement between network predictions and human vetters.
    }
    \label{fig:PredictionHist}
\end{figure*}

Fig. \ref{fig:PredictionHist} shows a histogram of network probabilities received by candidates for the \ortwo\ model. 
The fraction of candidates in a given bin which have been flagged either AS, BS, AD or D, is indicated by the colourbar.
As can be seen, candidates typically receive either low or high probabilities, with few clustered around 0.5. 
The vast majority of candidates receive a low probability from the network, consistent with the high false-positive rate previously established. 
Higher probability bins contain an increasing fraction of promising candidates, with AS, BS, AD or D flags, indicating a good general agreement between the neural network and the model.

While it is desirable to remove a large number of the false positives, caution needs to be taken not to exclude genuine planets from consideration. 
Fortunately in this case, approximately 50\% of candidates can be excluded using a conservative probability threshold of 0.1, reducing the time required to vet NGTS candidates by half.
We note that \citet{Osborn2019} and \citet{Dattilo2019} also favoured a threshold of 0.1.

From Table \ref{tab:planet_probs} it can be seen that the \ortwo\ model recovers the largest number of confirmed planets from models containing OFPs.
Similarly from Table \ref{tab:FlagPerformance} it is clear that this model also has the highest agreement fraction with eyeballing labels, among the models which include OFPs.
In deploying PlaNET as part of the NGTS pipeline we would like to be conservative, minimising the risk that promising candidates may be missed while accepting a slightly higher number of false positives.
Therefore we determine that our \ortwo\ model provides the optimum balance.
This is the only model which recovers all known planets, when using a threshold of 0.1, while still rejecting a substantial proportion of false positives.
It could be argued that since the \orone\ has the best overall AUC, considering a lower probability threshold may improve the recovery of candidates and outperform the \ortwo\ model. 
However in practise, even using a threshold of 0.1, two known planets would have been missed by the \orone\ model.

\section{New Candidates}
\label{sec:NewCands}

\begin{figure*}
	\centering
	\includegraphics[width=0.95\textwidth]{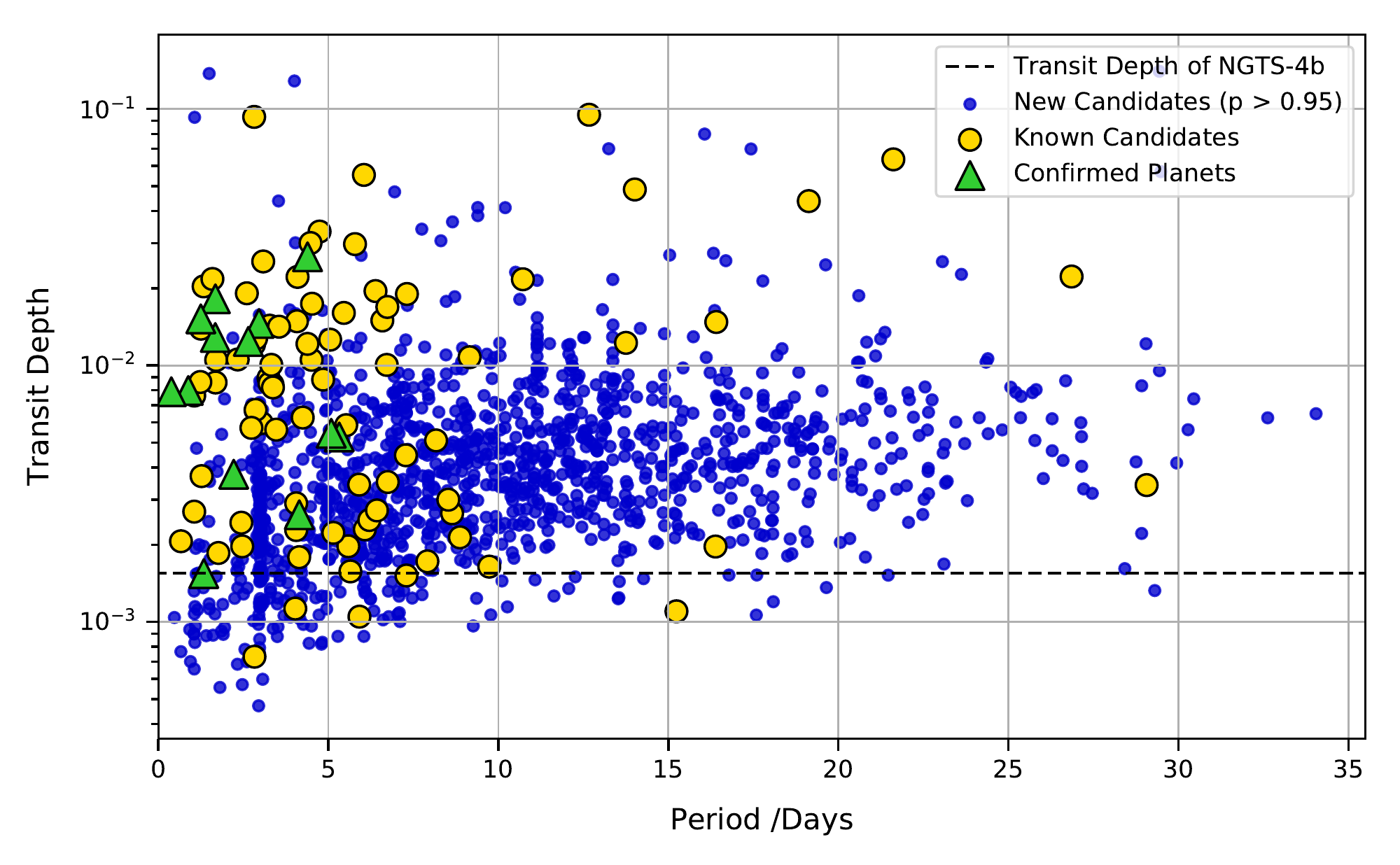}
    \caption{Transit depth versus orbital period for \textsc{ORION} candidates. For targets with more than one candidate detection, we adopt the detection with the highest network probability. Blue data points show `new' candidates, i.e. those with no eyeballing flags but with network probabilities greater than 0.95. Previously known candidates (with AS, BS and AD flags) are indicated by yellow data points where as confirmed planets are represented by green triangles. The black dashed line indicates the transit depth of NGTS-4b, currently the exoplanet with the shallowest depth detected from the ground in a wide field transit survey. There is no significant difference in the average depth of the different data series. Known candidates and confirmed planets typically have periods less than 5 and 10 days respectively, whereas new candidates span continuously up to periods of 35 days.}
    \label{fig:NewCand}
\end{figure*}

We used PlaNET trained on the \ortwo\ Max dataset, chosen as it had the highest AUC value, to identify new highly ranked candidates which had not previously been flagged by our vetters.
There are 13,253 such candidates with probabilities greater than 0.5, of which 1,309 have probabilities greater than 0.95. 

Fig. \ref{fig:NewCand} shows the transit depth versus orbital period for new candidates with probability greater than 0.95, compared with known candidates and confirmed planets. 
In general, transit signals with shallower depths are detected towards shorter orbital periods. 
This is likely because shorter periods allow a greater number of individual transits to be observed during the observing season, thus increasing the S/N of the transit in the phase folded light curves.

Transit depths for new candidates are strongly clustered around the 3 mmag level, which is comparable to known candidates and confirmed planets. 
Although the majority of known and confirmed planets lie at shorter orbital periods ($<$ 10 days), the period distribution of new candidates is broader, spanning up to 35 days. 
With fewer individual transits, these larger period signals are more susceptible to, and likely originate from artefacts in the light curves of individual nights. 
However, if validated they would increase the planet yield of the NGTS survey in this region of parameter space - since all currently confirmed planets have periods less than 5 days.

The network is not noticeably dissuaded from assigning high probabilities to large orbital period candidates. 
However, there is an apparent favouring of candidates with periods around 3 days for all depths. 
Since NGTS is a ground-based facility \textsc{ORION} ignores signals with periods within 5\% of 0.5, 1.0 and 2.0 days, where signals typically arise due to systematics strongly correlated with one sidereal day. This clustering at 3.0 days is also likely to be a one sidereal day alias.

\begin{figure*}
	\centering
	\includegraphics[width=0.95\textwidth]{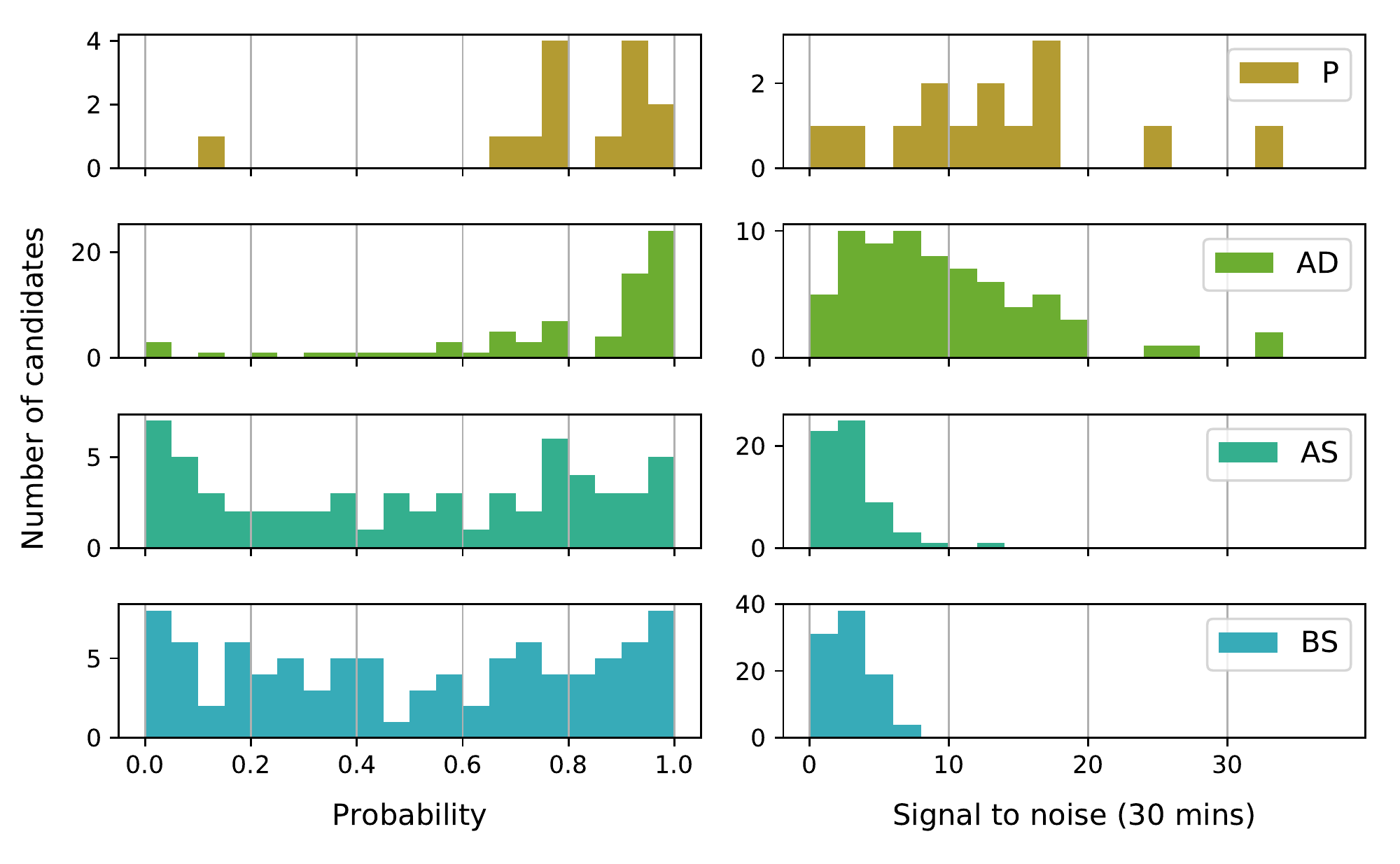}
    \caption{\textit{Left-hand panels:} Comparison of network probability distributions for candidates with P, AD, AS and BS flags (top to bottom), using the \ortwo\ Max model. Probabilities are the mean values averaged over 10 independent models, each trained with different portions of the overall dataset and different random seeds.  \textit{Right-hand panels:} Distribution in signal to noise for the same flags. Signal to noise is measured as per Fig. \ref{fig:DetectFrac}. On average, P and AD and flagged candidates have a higher signal to noise, with the probability distribution clustered towards larger values. This is consistent with a larger agreement fraction between the network and these flags, shown in Table \ref{tab:FlagPerformance}. The probability distribution for AS and BS flags is more uniform by comparison.}
    \label{fig:MeanProb}
\end{figure*}

When considering all candidates with probabilities greater than 0.5, we find that the vast majority of new candidates have low SDE. 
This is not surprising for several reasons. 
Firstly, the underlying distribution of \textsc{ORION} candidates is heavily skewed towards the low SDE range. With such a large number of \textsc{ORION} candidates being analysed by the network, a random subset of the new candidates will actually be false positives, but receive high probabilities due to statistical effects.
Therefore it is more likely these statistical false positives will have low SDEs.
Fig. \ref{fig:MeanProb} shows the probabilities for NGTS candidates, plotted with respect to the signal to noise of the detection.
The confirmed planets and AD candidates have higher S/N values, calculated as the transit depth divided by the standard deviation of the lightcurve, when phase folded and binned to 30 mins. 
The distribution of probabilities is split, with fewer in the range of 0.3 to 0.7, while the corresponding AS and BS distributions are much more uniform. 
This suggests that PlaNET is less certain about the nature of signals with lower S/N. 
It is also consistent with the lower accuracy in the selection of AS, BS candidates compared to AD candidates in Table \ref{tab:FlagPerformance}. 

Of additional consideration is that transit-like signals with higher SDEs are more easily identified during the vetting process, as they stand out more against the background noise. 
This is further reinforced by the fact that \textsc{ORION} candidates are presented in descending order of SDE and the vetter may become fatigued towards the bottom of the list. 
It is therefore more likely that overlooked candidates, will have low SDE. 
Similarly, low SDE candidates are less likely to be flagged during the vetting process as they are more ambiguous, more difficult to validate and their true nature is more likely to attract disagreement.

Though we expect most of these candidates to be false positives, this reinforces the point that the new candidates need to be carefully examined. Vetting and follow-up is on-going.

\section{Discussion}
\label{sec:Discussion}

We trained a convolutional neural network, called `PlaNET', to rank the 212,000 transiting exoplanet candidates identified in NGTS lightcurves. The network outputs a probability prediction of each candidate being an exoplanet. 
The main network inputs are the phase-folded NGTS lightcurves, but we also include inputs suggested from previous studies \citep{Ansdell2018,Osborn2019,Dattilo2019,Yu2019} which have been shown to increase performance.
Our motivation was to aid the manual candidate vetting process, by harnessing both the efficiency and consistency of a deep learning method. 
In doing so, we demonstrate that a large number of false positive candidates can be de-prioritised, depending on the choice of probability threshold. 
Even with a conservative threshold of 0.1, the network enables the confirmation effort to focus on the most promising 50\% of candidates, effectively reducing the vetting time by a factor of two. 

In this work, we focus on characterising how varying the network training dataset affects performance. 
Previous work has relied on the use of confirmed planets, as well as promising and rejected candidates determined via the vetting process, for their training dataset. 
In contrast, we also utilise injections of artificial planetary transits and false positive signals. For the non-planetary class, we consider various combinations of 4 false positive categories: (1) false positive candidates determined via vetting (OFP), (2) injections of stellar binary eclipses (EB), (3) lightcurves with no strong, periodic transit signals (NP) and, (4) transit and eclipsing binary signals folded on the wrong period (WP).

We validate the network's predictions by showing good agreement with candidate labels assigned by human vetters, as well as successful recovery of all but one of the 14 confirmed planet with NGTS lightcurves. 
Performance is particularly strong for deep transits and eclipsing binaries when both primary and secondary eclipse signals are clearly visible. 
Network models trained without OFPs in their datasets, recover all the confirmed planets. 
However, we find that the more OFPs included in the dataset, the more confirmed planets the network fails to recover, particularly for planets with higher signal to noise transits. 
A comparison of 4 different selection methods for inclusion of OFPs in the training data, showed that preferentially choosing the highest SDE OFPs gives better performance. This is as opposed to selecting OFPs randomly, uniformly or preferentially selecting those with the lowest SDEs.

Our results show that models trained using all 4 categories of false positives in the non-planetary class, perform almost as well as models trained solely on OFPs in this class; they achieve AUC values of approximately $76.5\%$ and $77.9\%$ respectively when measured on vetting labels. 
This suggests that in future, larger training datasets can be obtained by virtue of reduced reliance on labelled candidates from the vetting process. 
Our model of choice, \ortwo, achieves an AUC, accuracy, precision and recall of: $(76.5\pm{0.4})\%$, $(74.6\pm{1.1})\%$, $(0.98\pm{0.02})\%$ and $(63.0\pm{2.0})\%$ respectively on vetting labels.

Previous studies \citep{Pearson2018,Zucker2018,Osborn2019} explored the use of simulated data to train their networks. 
We present the first study which directly compares performance when training on fully simulated lightcurves versus real lightcurves with simulated planetary transits and eclipsing binaries, to compare how the noise properties of the data affects network performance. 
Although the network trained on fully simulated data performs best when validated on a test set of similar composition, the network trained using real data scores highest when assessing performance on the sample of NGTS lightcurves with vetting labels. 
This highlights that while fully simulated data allows the creation of larger datasets, adequately replicating the intricate noise properties of the real data remains an issue.

In addition, by utilising simulated data we present the first study of a CNN applied to transit lightcurves, which explores 2 important aspects of CNN training. 
Firstly, how network performance scales as a function of the number of lightcurves in the training dataset. 
Secondly, how performance is affected when training on lightcurves with incorrect labels. 
We find that additional gains in performance can be achieved by utilising larger datasets, beyond the sizes explored in both this work and previous work. 
As our results indicate, utilising transit injections and incorporating additional categories of false positives appears to be a viable way of expanding the dataset to increase network performance. Incorrect lightcurve labels may arise for several reasons, particularly for genuine, low signal to noise transits which are not identified in the vetting process. 
It is easy to see how this might confuse the network while it is learning, an issue discussed by \citet{Zucker2018} and \citet{Yip2019}. 
Knowledge of the ground truth is one of the main advantages of training on simulated data. 
Interestingly however, we find that our networks are robust to contaminated labels; only minor degradation in overall performance is experienced up to a contamination fraction of 0.48, after which performance decreases rapidly. 
This result is consistent with those from other studies \citep{Rolnick2017,Li2019,Reis2019} and suggests that label contamination in real data is of little consequence to overall performance.

Finally, our analysis identified `new', highly ranked candidates which had not previously been flagged by the NGTS team. 
There are 13,253 such candidates with probabilities greater than 0.5, of which 1,309 have probabilities greater than 0.95. 
At the time of writing, further scrutiny of these new candidates is ongoing. 
Interestingly, the period distribution of these candidates extends continuously up to 35 days, whereas previously known NGTS candidates and confirmed planets lie predominantly below 10 days and 5 days respectively. 
While likely to be false positives, if any of these new candidates are confirmed, they may present an opportunity to substantially expand the parameter space in which NGTS is finding planets.
\\\\
\noindent We highlight several areas of improvement for future work:
\begin{itemize}[itemsep=1em,leftmargin=1em]
\item Our networks do not recover all the confirmed planets nor all the high S/N transits, particularly when there are more OFPs in their training dataset. On test data, the \orone\ models recover the most high S/N candidates. While comparing against NGTS vetting labels, the \ortwo\ Max recovers the most deep transit candidates.
This difference was consistent across the entire ensemble trained for each model. Further work is needed to clarify why exactly this is happening, though we have two main hypotheses. This may be because there are similar signals in the non-planet class of training data, which cause the network to favour a non-planet classification in these cases. Or alternatively, although we carefully sampled period and stellar radius parameters to reduce network bias between the planet and non-planet classes, we made no attempts to reduce bias within each class, with respect to parameter distributions such as the S/N. Our Monte Carlo injection method exacerbated this issue by producing non-uniform posterior distributions. Ideally we would prefer to construct datasets with more uniform parameter distributions, though this is difficult to accomplish since there are many parameters with complex inter-dependencies, and we would be limited by the number of bright targets in our data. Finally, we could employ the use of tools to gain additional insight into the network's logic behind mis-classifications \citep{Philbrick2018}, such as class activation maps \citep{Zhou2015} used in \citep{Yip2019}, and visualisations of the final hidden layer geometric space in fewer dimensions \citep{vanDerMaaten2008}.
\item We showed that using a larger training dataset yields better results. When training on real data, we used a total of 24,000 lightcurves for all models. This choice was a practical compromise between maximising performance and minimising the time for data generation and preparation. However if we utilise all available data, we estimate that the training dataset could be nearly doubled to 41,000 lightcurves, assuming no inputs are rejected by our bad data filtration criteria. The main limitation to the dataset size comes from the OFP model, specifically the number of false positive candidates identified by \textsc{ORION}. If instead, we consider only models with more than one category in the non-planet class, we can increase the training dataset size further. We showed that the \ortwo\ model was actually better overall for planet recovery than the OFP model, and our preferred choice for the deployment of PlaNET in the NGTS pipeline.
\item A prevailing trend across previous applications of CNNs to transit lightcurve classification, is that adding additional network inputs tends to increase performance. Increasing the number of auxiliary scalar parameters is trivial since choices are in abundance and they have minimal impact on computation time. \citet{Osborn2019} utilised 16 auxiliary scalar parameters, mostly associated with stellar parameters, however in this work we considered only three. This decision was motivated mainly by our use of simulated data, for which producing a self-consistent set of additional stellar parameters is non-trivial. However, if we were to consider only real data, then we could expand the number of parameters.
\item We assessed network performance using the NGTS database of candidate labels, assigned during the main consortium vetting process. As such it is likely that our network performance was lower with respect to the human vetters, since NGTS eyeballers had access to additional information at the time of making their assessment, which the network did not. For instance: follow-up photometry, radial velocities, results of fitting - all which can change the outcome completely. In contrast, \citet{Yu2019} 
carried out their own labelling exercise specifically for the network; conducting a similar process for NGTS would increase the reliability of our results. 
\item We would like to make a detailed comparison of the performance of PlaNET to the Autovetter \citep{Armstrong2018} tool. Any systematic differences between the two algorithms may highlight ways the design of PlaNET can be improved and which additional information could be included to boost performance, e.g. stellar parameters, transit information, etc.
\item We conducted a limited study to optimise our network hyperparameters. We found no statistically significant combination of hyperparameters which maximised performance. For lack of a better choice, we adopted the same network architecture as \citet{Shallue2018}, with differences in: batch size, number of epochs, dropout probability and the local view time span. Unlike Kepler, NGTS is a ground-based instrument with completely different noise properties; there is no evidence to indicate that the Shallue architecture is also optimal for NGTS lightcurves. A complete optimisation using traditional grid or Baysian TPE methods would have been prohibitively expensive. We note that the majority of similar studies also carried out limited optimisation exercises. Nevertheless, alternative methods for optimising neural architecture could be investigated.
\end{itemize}

\section*{Acknowledgements}

Based on data collected under the NGTS project at the
ESO La Silla Paranal Observatory. The NGTS facility is
operated by the consortium institutes with support from
the UK Science and Technology Facilities Council (STFC)
through projects ST/M001962/1 and ST/S002642/1. LR is supported by an STFC studentship (1795021). The contributions at the University
of Leicester by MRG and MRB have been supported by
STFC through consolidated grant ST/N000757/1. PE and ACh
acknowledge the support of the DFG priority program SPP
1992 "Exploring the Diversity of Extrasolar Planets" (RA
714/13-1). 
The contributions at the University of Warwick by PJW and RGW
have been supported by STFC through consolidated grants ST/L000733/1 and ST/P000495/1. DJA gratefully acknowledges support from the STFC via an Ernest Rutherford Fellowship (ST/R00384X/1). 
JSJ acknowledges support by Fondecyt grant 1161218 and partial support by CATA-Basal (PB06, CONICYT).
This work has made use of data from the European Space Agency
(ESA) mission Gaia (https://www.cosmos.esa.int/gaia),
processed by the Gaia Data Processing and Analysis Consortium (DPAC, https://www.cosmos.esa.int/web/gaia/dpac/consortium). Funding for the DPAC has been provided by national institutions, in particular the institutions participating in the Gaia Multilateral Agreement.
This research has made use of the NASA Exoplanet Archive, which is operated by the California Institute of Technology, under contract with the National Aeronautics and Space Administration under the Exoplanet Exploration Program.
This research used the ALICE High Performance Computing Facility at the University of Leicester.




\bibliographystyle{mnras}
\bibliography{references} 







\bsp	
\label{lastpage}
\end{document}